\documentclass[reprint,amssymb,amsmath,aip,cha]{revtex4-1}
\usepackage{graphicx}
\usepackage[caption = false]{subfig}
\usepackage{color}
\usepackage{epsfig}
\usepackage{ifpdf}
\usepackage{bm}
\usepackage[colorlinks=true,linkcolor=blue]{hyperref}%
\expandafter\ifx\csname package@font\endcsname\relax\else
 \expandafter\expandafter
 \expandafter\usepackage
 \expandafter\expandafter
 \expandafter{\csname package@font\endcsname}%
\fi
\usepackage{cleveref}
\begin{document}
\title{Comparative Study of the Surface Potential of Magnetic and Non-magnetic Spherical Objects in a Magnetized RF Discharge Plasma}
\author{Mangilal Choudhary}
\affiliation{I. Physikalisches Institut, Justus--Liebig Universität Giessen, Henrich--Buff--Ring 16, D 35392 Giessen, Germany}
\author{Roman Bergert}
\author{Slobodan Mitic}
\author{Markus H. Thoma}
%
\begin{abstract}
 We report measurements of the time-averaged surface floating potential of magnetic and non-magnetic spherical probes (or large dust particles) immersed in a magnetized capacitively coupled discharge. In this study, the size of the spherical probes is taken greater than the Debye length. The surface potential of a spherical probe first increases, i.e. becomes more negative at low magnetic field (B $<$ 0.05 T), attains a maximum value and decreases with further increase of the magnetic field strength (B $>$ 0.05 T). The rate of change of the surface potential in the presence of a B-field mainly depends on the background plasma and types of material of the objects. 
The results show that the surface potential of the magnetic sphere is higher (more negative)  compared to the non-magnetic spherical probe. Hence, the smaller magnetic sphere collects more negative charges on its surface than a bigger non-magnetic sphere in a magnetized plasma. The different sized spherical probes have nearly the same surface potential above a threshold magnetic field (B $>$ 0.03 T), implicating a smaller role of size dependence on the surface potential of spherical objects. The variation of the surface potential of the spherical probes is understood on the basis of a modification of the collection currents to their surface due to charge confinement and cross-field diffusion in the presence of an external magnetic field. 
\end{abstract}  
\maketitle
\section{Introduction}
A spherical object or dust grain attains an equilibrium potential (or floating potential) when it is immersed in a plasma. At the floating potential, it draws a net zero current, i.e., the net flux of electrons and ions to the surface of the spherical object is zero. In low-temperature plasmas, where the electron temperature is much higher than the ion temperature ($T_e \gg T_i$), the floating potential of the spherical particle mainly depends on the flux of energetic electrons to its surface and is always negative with respect to the plasma potential. In a dusty plasma, which is an admixtures of the plasma species and sub-micron to micron sized solid particles, the charge on the dust grains determines their collective dynamics such as dust acoustic waves \cite{daw2,pdasw,mangilaldaw,vikramtsw}
and vortex motion \cite{Vaulina2004,satoprl,mangilalmultiplevortex,probeinducedcirculation, mangilallargeaspect,mangimagneticrotation}.
In a dusty plasma, a dust particles are assumed to be spherical capacitors, which allows us to  determine the surface potential and the net charge on it.\par 
In recent years, the research field of dusty or complex plasmas has been of interest due to its applications in space or solar plasmas, \cite{goertzdustysolarsystem,geortzspokes2,cosmicdustymendis} plasma processing technologies, \cite{selwynprocessing1,watanableprocessingplasma2} fusion devices, \citep{winterfusiondust} colloidal solutions \cite{colloidaldustyplasma} etc. For studying the collective dynamics of the dust grain medium, the charge on the dust grains has to be known. In the last more than three decades, various experimental methods have been used to obtain the dust charge in an unmagnetized dusty plasma. \cite{Charging,goreedustcharging1,jwangdustcharge1,walchdustcharge2,particlechargemarkus1} The experimentally measured dust charge values were compared with theoretically obtained values using the orbital-motion-limited (OML) approximation \cite{mottsmitomltheory1,allenomltheory2} and numerical simulations \cite{particlechargemarkus1}. The OML theory describes the charging mechanism of sub-micron to micron-sized particles ($r <\lambda_{De}$) in the plasma environment. Here, $r$ is the radius of the particle and $\lambda_{De}$ is the electron Debye length. For large dust grains or spherical objects ($r > \lambda_{De}$), the thin sheath theory or the modified OML theory \cite{willisfloatingpotential2} is suitable to understand the charging mechanism in an unmagnetized plasma.\par
Nowadays, magnetized dusty plasma are a  popular research topic among the dusty plasma community. It is well known that the dynamics of dusty plasmas depends on the characteristics of the background plasma that can be changed in the presence of an external magnetic field. Therefore, the B-field is considered as an external parameter to control the dynamics of the dust grain medium which may allow to us study fluid dynamics, solid state phenomena, turbulence etc. at a microscopic level. \cite{morfilldusty1,bonitzdusty2}
Since the dust charge depends on the background plasma, the estimation of the accurate charge on dust grains (magnetic or non-magnetic) is a challenge depending on the magnetization of the plasma particles. In the last few years, theoretical as well as experimental works have been carried out to estimate the charge on dust grains in the magnetized plasma. Tsytovich \text{et al.} \cite{tsytovichdustcharge1} performed simulations to understand the charging mechanism of micron sized dust particles in a magnetized plasma. It has been claimed that the B-field influences the dust charging mechanism at a strong B-field when the electron gyration radius is greater than the dust radius. Lange \cite{langefloatinginmagnetized} performed a simulation of a magnetized rf plasma and observed a smaller dust surface potential (or charge) at a lower magnetic field. The simulation of Patacchini \textit{et al.} \cite{dustcurrent} demonstrated the decrease of the dust charge at all values of magnetic field in a collisonless plasma. A recent simulations \cite{kodanovadustchargeb} suggests that the dust charge starts to decrease after a critical value of B-field in a magnetized plasma. Tomita \textit{et al.}\cite{tomitadustchargingwithmagneticfield} reported a higher dust surface potential or large dust charge in a weakly magnetized plasma. Apart from analytical and numerical simulation studies, a few experiments have been performed to obtain the dust charge in a magnetized plasma. Kalita \textit{et al.} \cite{kalita} have measured the dust charge in a weakly magnetized plasma (B $<$ 0.05 T) and found a negligible role of  the B-field to be negligible on the dust charging mechanism. Tadsen \textit{et al.} \cite{melzermagnetizeddusty} have observed a reduction of the dust charge up to 50\% for nano sized particles at a low magnetic field (B $<$ 0.01 T) in an rf discharge. In this work, the charge on nano sized non-magnetic particles is determined by fitting the theoretical dispersion relation of dust--acoustic waves to the experimentally observed dispersion relation. An experimental work of Melzer \textit{et al.} \cite{melzerdustchargeb} shows a reduction in the dust charge at low magnetic field (B $<$ 0.02 T) and nearly constant value up to B $>$ 5 T. In their work, the charge on micron sized dust grains is extracted by a normal mode analysis of the dust cluster in the magnetized rf plasma. A similar rf discharge configuration and normal mode analysis technique were used to get the charge on micron sized paramagnetic particles at low B-field (B $<$ 0.01 T) and observed a nearly constant values of the dust charge. \cite{paramagneticdust} 
The inconsistencies in the numerically as well as experimentally observed values of the dust charge leave many open questions on the charging mechanism of spherical particles (magnetic and non-magnetic) in a magnetized plasma. Does the dust charge remain unchanged even though the magnetic field modifies the background plasma? How does it depend on the density of the plasma species and the background gas in the presence of B-field? Do magnetic particles attain similar charges than non-magnetic particles? Does the surface potential of a dust grain show a size dependence in a magnetized plasma? Why does the experimentally estimated dust charge not vary according to the theoretical models? \par
 To answer some of these questions, a better understanding of the dust charging in an rf magnetized plasma is required. It is sometimes difficult to measure a small variation in the charge of micron-sized dust grains ($r < \lambda_{De} $) while the background plasma parameters are changing in the presence of an external magnetic field. In laboratory experiments, it is easy to directly measure the surface potential of a large spherical conducting body ($r > \lambda_{De} $), which can be considered as a large dust grain in a magnetized plasma. The surface potential variation of a spherical probe (or large dust grain) in presence of a external magnetic field can provide information of a background plasma to minimize the errors in measuring the charge on micron-sized particles ($r < \lambda_{De} $) in a magnetized dusty plasma. It is well known that the charge on an individual dust particle is higher (more negative) than that of a particle in the dust grain cluster. However, the variation of the dust charge of a single dust grain and dust cluster would be similar. Sometimes the surface potential of a large dust grain also helps to understand the interactions among the micron-sized dust grains in a magnetized plasma. Keeping this in mind, experiments are planned to measure the surface potential of magnetic and non-magnetic spherical probes (or large dust particles) in a magnetized rf discharge.
\par
The investigations are carried out in a magnetized complex plasma device where an rf glow discharge is ignited between two electrodes, and a superconducting electromagnet is used to introduce the magnetic field. The surface potential of various sized magnetic (stainless steel, SS-430, $\mu_r$ = 1800) and non-magnetic (bronze, $\mu_r$ $\sim$ 1) spherical probes (or large dust grains) has been measured in the unmagnetized and magnetized plasma at various discharge conditions. At a lower magnetic field, the magnitude of the surface potential of a spherical object increases to a maximum value and then starts to decrease with increasing the strength of the B-field. This trend is found to be independent of the size and types of materials of the spherical object. However, the charging mechanism of magnetic and non-magnetic spherical objects depends on the magnetic field. The charge or surface potential of a non-magnetic spherical probe in the plasma are found to be smaller (less negative) than that of a  magnetic sphere after a threshold value of B-field. Experimentally observed results are explained on the basis of the current collection to the surface of the object in the presence of a magnetic field. \par
The manuscript is organized as follows: Section~\ref{sec:exp_setup} deals with the detailed description of the experimental set-up and the magnetized plasma production. The surface floating potential variation at various discharge conditions in unmagnetized and magnetized plasmas are discussed in Section~\ref{sec:results_Vf_B}. Qualitative and quantitative explanations of the surface potential variation for magnetic and non-magnetic spheres are given in Section~\ref{sec:results_discussion}. A brief summary of the work along with concluding remarks is provided in Section~\ref{sec:summary}.
 \section{Experimental Setup and Diagnostics} \label{sec:exp_setup}
\begin{figure*}
 \centering
\subfloat{{\includegraphics[scale=0.80]{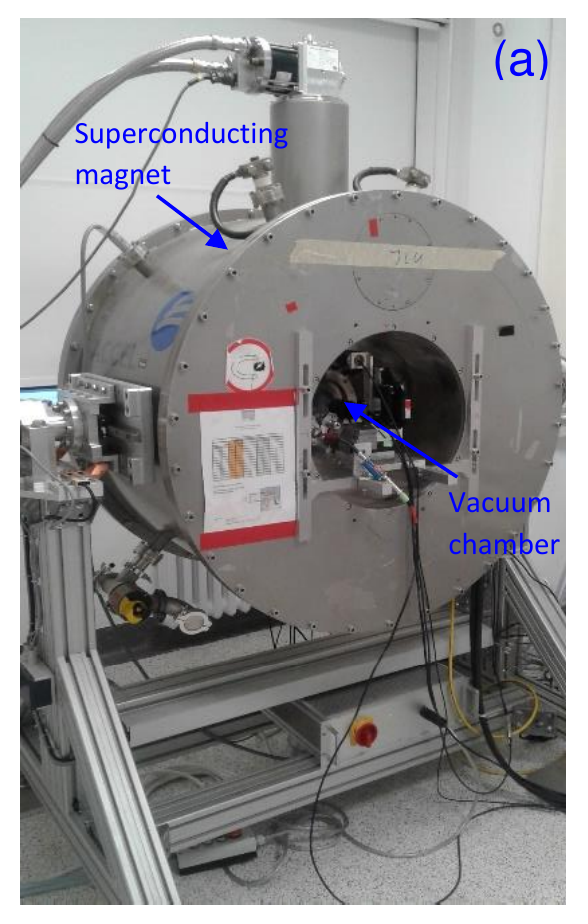}}}%
\hspace*{0.4in}
 \subfloat{{\includegraphics[scale=0.55150]{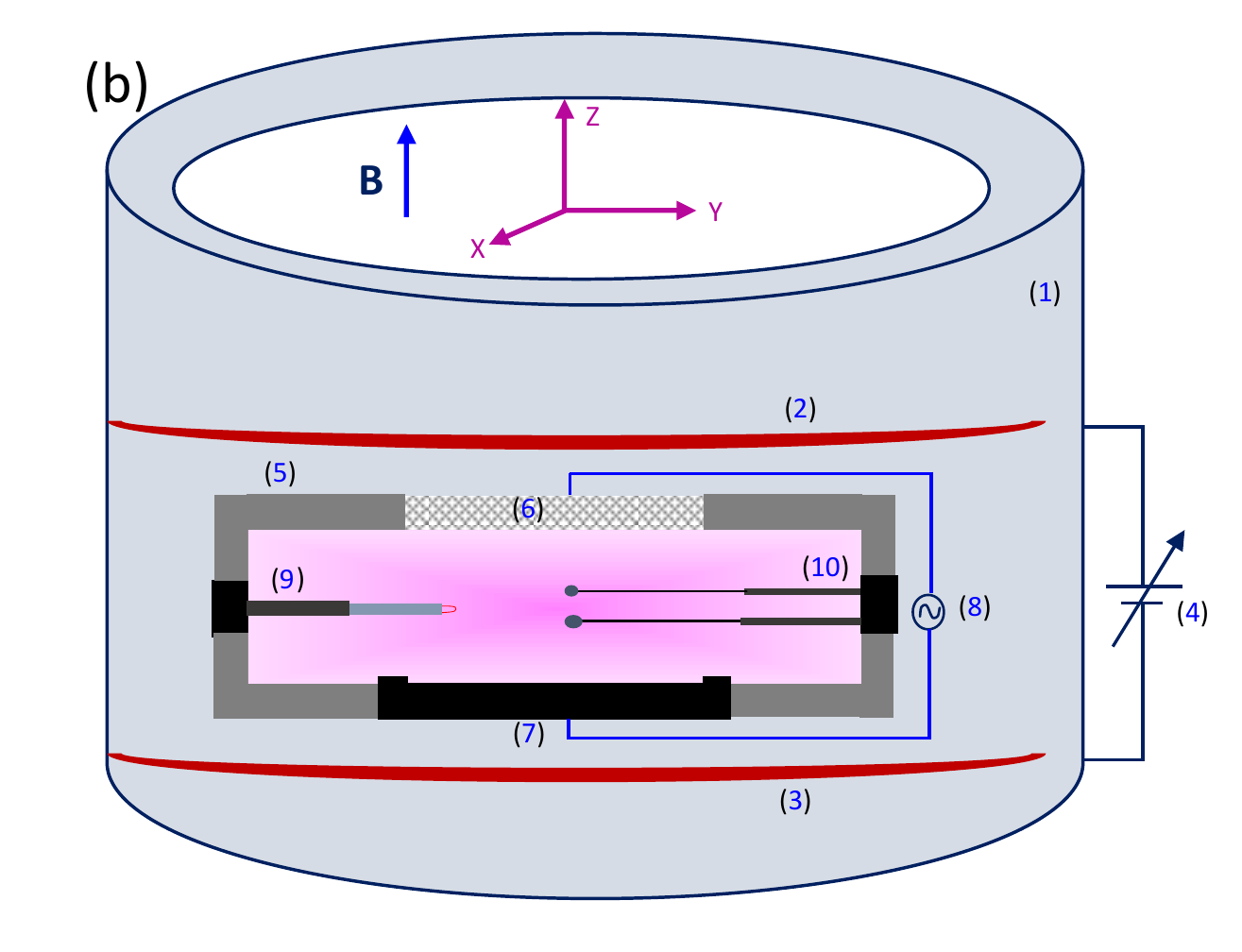}}}
 \qquad
 \subfloat{{\includegraphics[scale=0.75150]{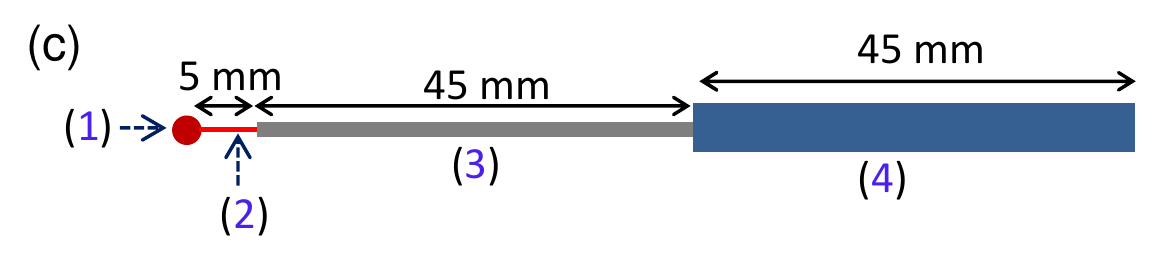}}}
 \caption{\label{fig:fig1}(a) Magnetized dusty plasma device. (b) Schematic diagram of the experiment setup (1) superconducting electromagnet, (2) and (3) are superconductor coils, (4) power supply for magnet, (5) vacuum chamber, (6) upper electrode, (7) lower electrode, (8) 13.56 MHz rf generator with matching network, (9) emissive probe and (10) spherical probes. The direction of the magnetic field along the Z-direction is represented by a blue arrow. (c) schematic diagram of the spherical probe (1) condcuting (non-conducting) sphere, (2) connecting wire (insulation shielded), (3) ceramic tube of diameter 2 mm, and (4) aluminium tube of diameter 6 mm. } 
 \end{figure*}
The experimental setup (magnetized dusty plasma device) consists of an aluminum vacuum chamber and a superconducting electromagnet, which is shown in Fig.~\ref{fig:fig1}(a). This device was previously used to study the dusty plasma in a strong magnetic field \cite{kanopkamagnetized}. The schematic diagram of the experimental setup is presented in Fig~.\ref{fig:fig1}(b). The superconducting electromagnet has a Helmholtz coil configuration to produce a uniform magnetic field at the center of the vacuum chamber. The superconducting magnet consists of a helium compressor, a cooling head, 8 sensors for temperature measurements, and a superconducting magnet power supply (0 to 80 A). At first, the plasma chamber is evacuated below $10^{-2}$ Pa using a pumping system consisting of a rotary and turbo molecular pump (TMP). The experiments are performed in an argon plasma and the pressure of the argon gas inside the chamber is controlled by a mass flow controller (MFC) and a gate valve controller. A 13.56 MHz rf generator with matching network is used to ignite the gas discharge between a stainless steel electrode (lower) and an indium tin oxide (ITO) coated glass electrode (upper) of  6.5 cm diameter. Both electrodes are separated by 3 cm. For the comparative study, stainless steel (SS-430, $\mu_r$ = 1800, magnetic) spherical probes of radius 1.0 mm, 1.25 mm and 1.7 mm and a bronze (non-magnetic, $\mu_r$ $\sim$ 1) spherical probe of radius 1.5 mm are used. Opposite radial ports are used to insert the spherical probes and the emissive probe for measuring the floating and plasma potentials. The measurements are taken in the homogeneous plasma region where the magnetic field is uniform. The spherical probes are placed in the plasma using a ceramic tube of diameter 2 mm which protrudes into the plasma by a feed-through in the chamber wall and holds the spherical probes at its end. 
To avoid the floating potential perturbation due to the connecting aluminium tube (it holds ceramic tube), the length of the ceramic tube is taken longer so that it only remains in contact with the plasma, as is shown in Fig.~\ref{fig:fig1}(c). We have made connection to the spherical probe in such a way to keep connection area as small as possible compared to the total surface area. For measuring the time-averaged floating potential ($V_f$) of a spherical probe (or large dust grain), a high-impedance voltage divider (1200:1) is used. The spherical probe is connected to a high-value resistor ($R_1$ = 120 M$\Omega$) to minimize the current flowing in the voltage divider circuit. First the voltage drop ($V_2$) with respect to ground due to this small current is measured across a low value resistor ($R_2$ = 100 k$\Omega$) and then the floating potential of the spherical probe ($V_f$) is calculated by using the expression, $V_f = (R_1 + R_2) V_2 /R_2 $. In the present set of experiments, an emissive probe made of tungsten of radius 0.05 mm, placed perpendicular to the magnetic field lines, is used to measure the time-averaged plasma potential ($V_p$). The floating point method technique is used to measure $V_p$ in the absence and presence of the  magnetic field\cite{emissivesheehan,balanemissiveprobetokamak,hitujaemmisiveprobe,bradelyemmisiveprobeinB,mangilalthesis}. 
 It has been claimed in some studies that the floating potential method underestimates the plasma potential \cite{emissiveprobescrutwiser,emissiveprobecomparisionsheehan}. In view of this, we have compared the plasma potential values obtained from the cold single Langmuir probe and emissive probe. The floating potential method estimates the plasma potential lower by $<$ 2 V than estimated by a cold probe. This potential difference is within the error of $<$ 15 \%. Thus, the floating potential method is useful to get an approximate value of plasma potential to obtain $V_s$.
\section{Measurements of surface potential of spherical objects} \label{sec:results_Vf_B}
A spherical object or dust grain immersed in a plasma attains a negative potential to balance both the electrons and ions currents to its surface. This equilibrium surface potential with respect to the plasma potential ($V_p$) is termed as surface potential ($V_s = V_p - V_f $) of the spherical object \cite{chenprobe1,sphericalprobe2}. It is stated in ref.\cite{willisfloatingpotential2} that different analytical theories are valid for estimating the surface potential of an object in a Maxwellian plasma. The orbital motion limited (OML) theory \citep{allenomltheory2} is applicable for small objects ($\rho = r/\lambda_{De} << 1$) and the surface potential is derived by balancing the electrons and ions fluxes to the surface of an object [ref.\cite{willisfloatingpotential2,beadlesfloatingpotential1}]
\begin{equation}
 exp(-\Phi)  =  \sqrt{\frac{T_i}{T_e} \frac{m_e}{M_i}}\bigg(1 - \frac{V_s}{T_i}\bigg) 
\end{equation}
where $\Phi = -V_s/T_e$, $T_e$ and $T_i$ are the electron and ion temperatures, $m_e$ and $M_i$ are the electron and ion masses, respectively. 
For the large spherical object ($\rho >> 1$), the thin sheath theory (TS) is applicable to obtain $V_s$ . The surface potential for such spherical objects can be estimated by \cite{thinsheath,willisfloatingpotential2}
\begin{equation}
 -V_s = \frac{T_e}{2} \bigg[\Big( 2 \pi \frac{m_e}{M_i}\Big)\Big( 1 + \frac{T_i}{T_e}\Big)\bigg]  + ln2
\end{equation}
In the transition region between OML and TS theory, the orbital motion (OM) theory estimates  the surface potential. Then $V_s$ is found to be a straight line fit on a log plot between the OML limit and  TS limit. \cite{beadlesfloatingpotential1,willisfloatingpotential2} It is clear from Eq. 2 and Eq.3 that the floating surface potential ($V_s$) of a large spherical object or small dust grain has a approximately linear relation with $T_e$,
 \begin{equation}
 V_s = - \alpha T_e ,
\end{equation} 
Here $\alpha$ is a constant varying from $\sim$ 2.5 to 4 in the transition region between the OML and TS  ($0.1 < \rho <10$)  for an unmagnetized argon plasma ($T_i << T_e$). \cite{willisfloatingpotential2,beadlesfloatingpotential1} In a magnetized plasma, $V_s$ also depends on $T_e$ but the value of $\alpha$ may be lower or higher than that of an unmagnetized plasma.
 For getting the value of the floating surface potential, $V_s = V_p - V_f$, of a spherical conducting probe in the plasma, it is necessary to measure the plasma potential ($V_p$) as a reference potential. By knowing the surface potential ($V_s$), the charge on the surface ($Q_s$) of a small dust grain ($r <\lambda_{De}$) and large dust grain ($r >\lambda_{De}$) can be estimated using the different approximations \cite{chargingsphere}.
\subsection{Surface potential of spherical probes in unmagnetized plasma}
The present work deals with spherical probes (large dust grains) of radius larger than the electron Debye length, i.e., $r > \lambda_{De}$. Stainless steel spheres of radius 1.0 mm, 1.25 mm and 1.7 mm are used to study the size dependence of the surface potential. A pair of spherical probes of different sizes (separated by 14 mm) is placed in the plasma volume, as shown in schematic diagram (see Fig.~\ref{fig:fig1}(b)). It should be noted that both probes are placed in the horizontal (X--Y) plane. The distance between the probes is decided after successive measurements of $V_f$ for both spherical probes at similar discharge conditions in the presence of B-field. These successive measurements are taken at the center of the plasma volume whereas the simultaneous measurements on both sides of the center are performed to keep both probes in the homogeneous plasma background. The difference between the successive and simultaneous measured values of $V_f$ at the same discharge condition are found to be $<$ 0.3 V, which is $<$ 2--3 \% of the actual value. Therefore, we neglect the shadow/potential overlapping effect of an individual sphere on each other during the simultaneous measurements of $V_f$ for the comparative study. It is known that dust grains ($\mu$m to mm) respond only to a very low frequency external field ($\sim$ 1 to 100 Hz). They do not respond to a high frequency field of an rf discharge. In view of this, it is our aim to measure the time-averaged or DC potential of the spherical probes in the rf discharge.
\par
\begin{figure*}
 \centering
\subfloat{{\includegraphics[scale=0.30]{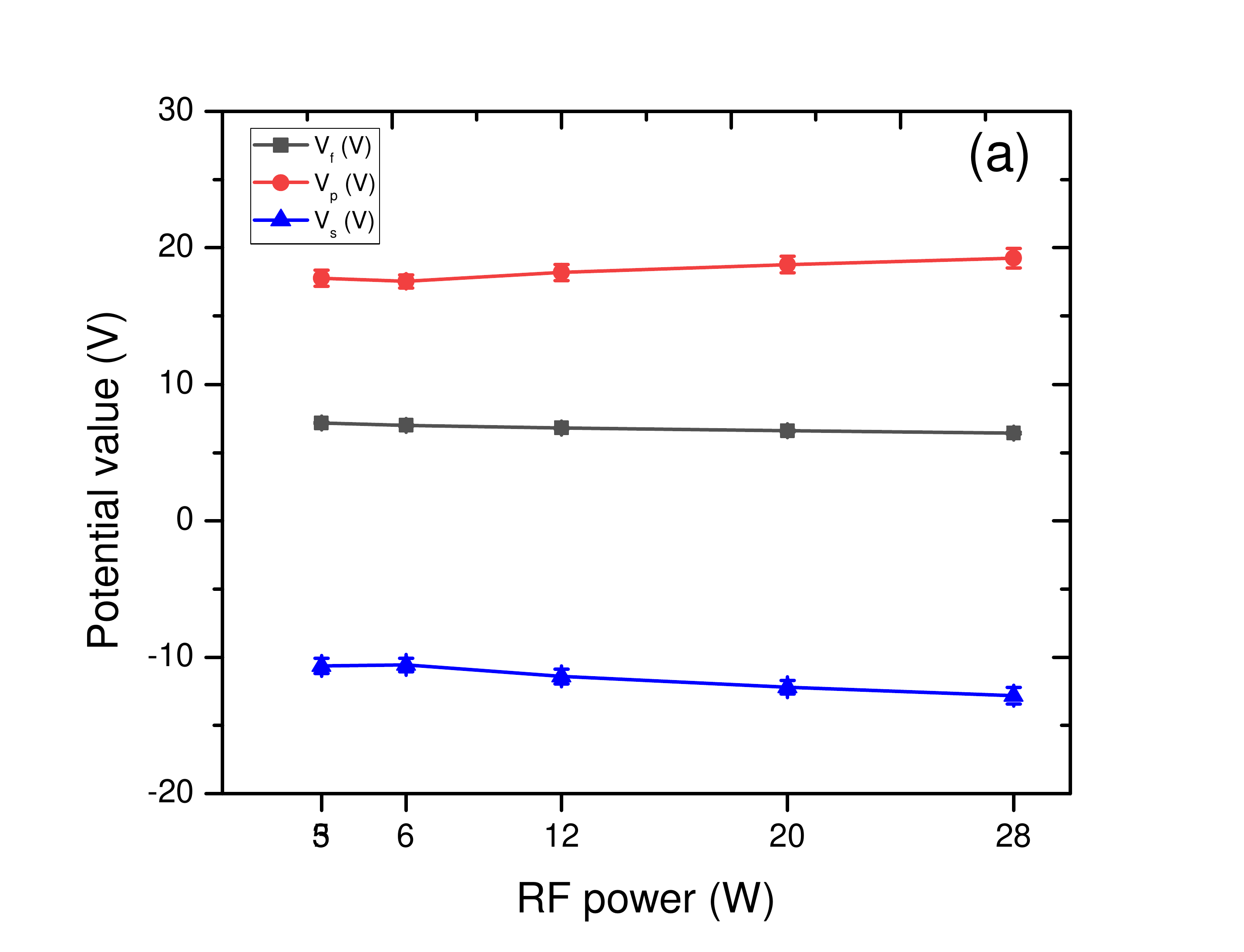}}}%
 \subfloat{{\includegraphics[scale=0.300]{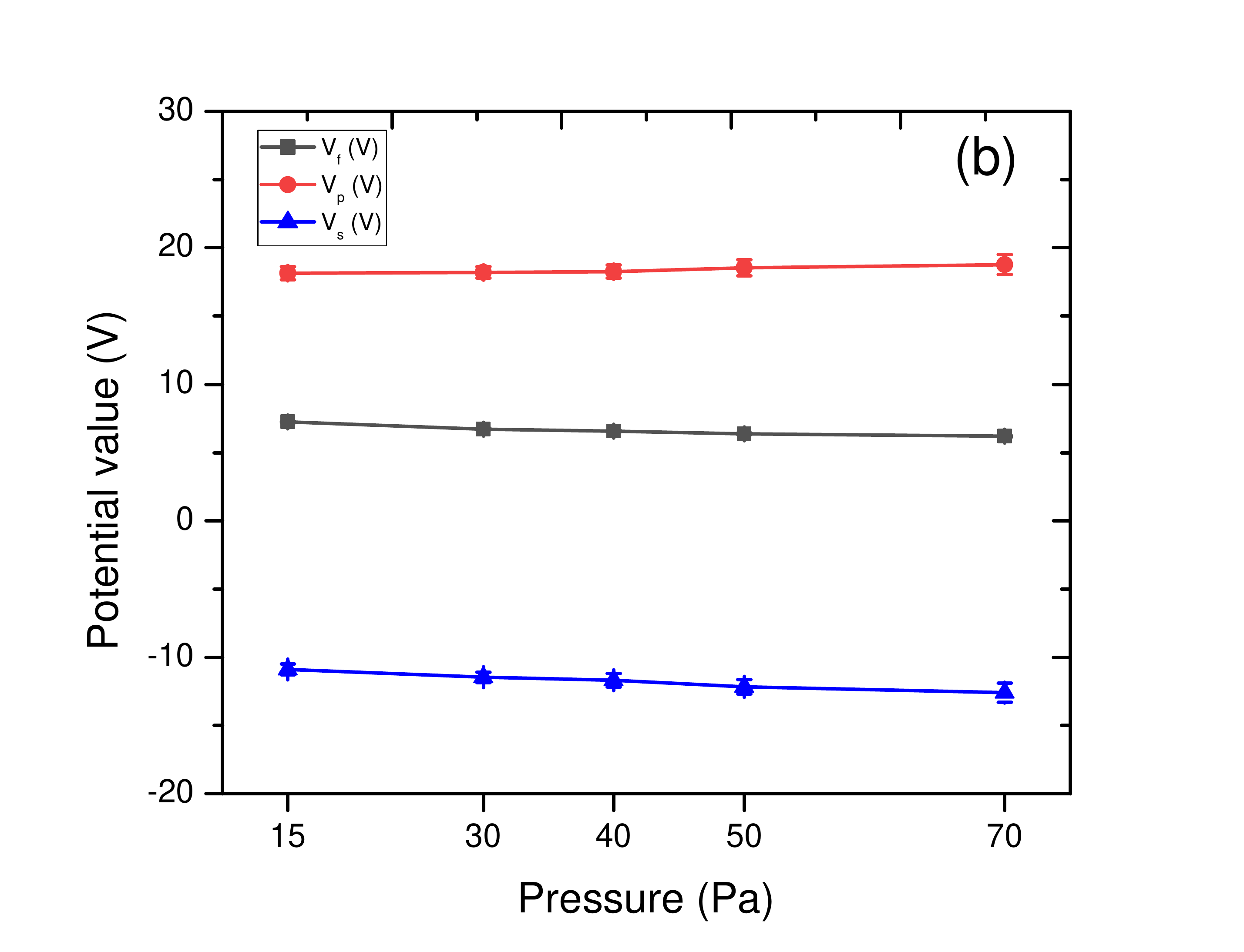}}}
 \qquad
 \caption{\label{fig:fig2} (a) Time-averaged floating potential ($V_f$), plasma potential ($V_p$) and surface potential ($V_s$) of the spherical stainless steel probe ($r$ = 1.7 mm) for different input rf powers at fixed pressure, p = 30 Pa in an unmagnetized plasma. (b) $V_f$, $V_p$ and $V_s$ of the same spherical probe for different argon pressures at fixed rf power P = 12 W in an unmagnetized plasma. The plotted values of $V_f$ and $V_p$ are averaged over few data sets at given discharge condition.} 
 \end{figure*}
\begin{figure}
\centering
 \includegraphics[scale= 0.3000]{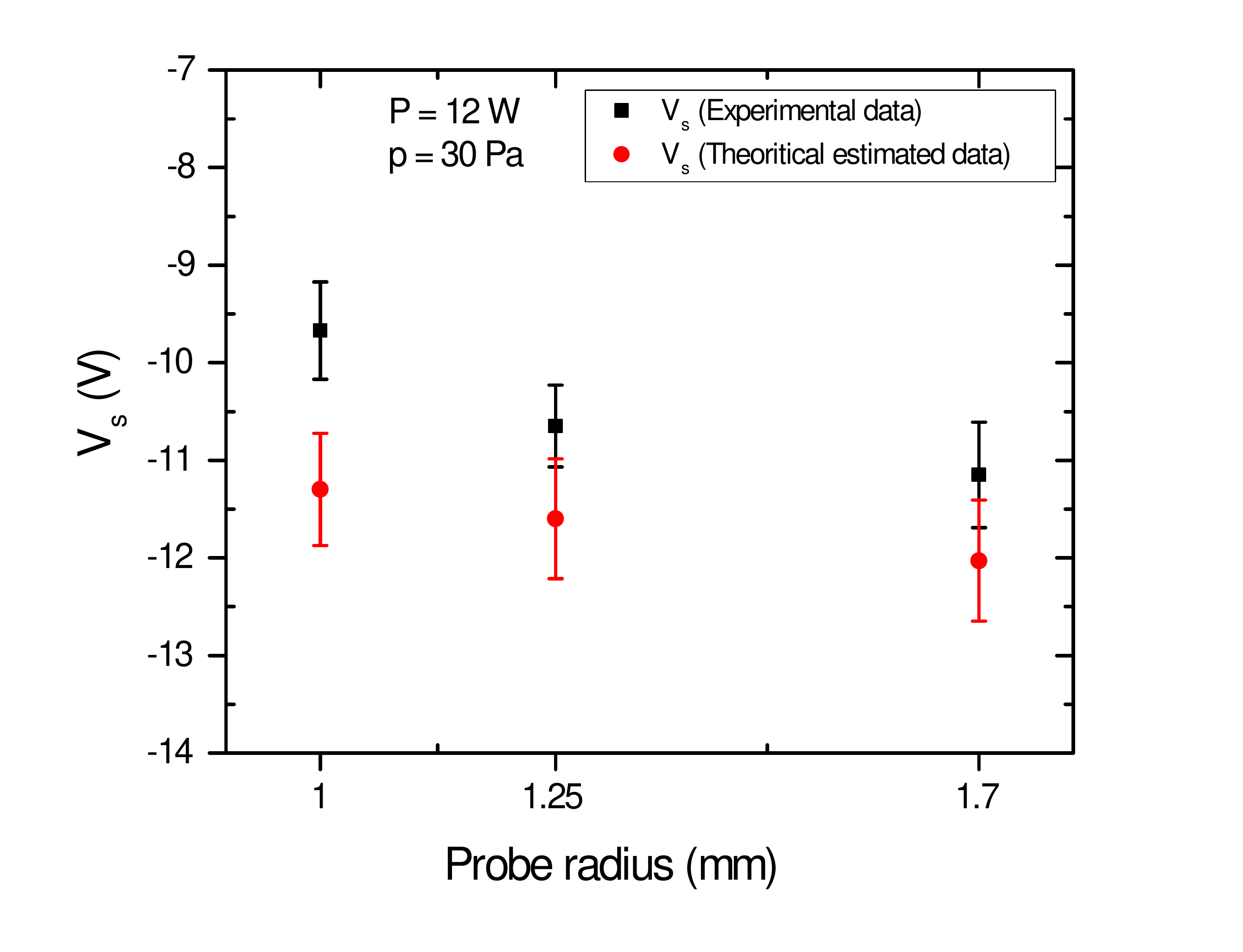}
\caption{\label{fig:fig3} Experimental and theoretical surface potentials ($V_s$) of different sized stainless steel spherical probes ($r$ = 1, 1.25 and 1.7 mm) at rf power, P = 12 W and gas pressure, p = 30 Pa in an unmagnetized plasma. The error in measuring the plasma potential is $< \pm$ 5 \% and floating potential is $< \pm$ 2 \%. Both errors are included over the average values.}
\end{figure}
\begin{figure}
 \includegraphics[scale= 0.3000]{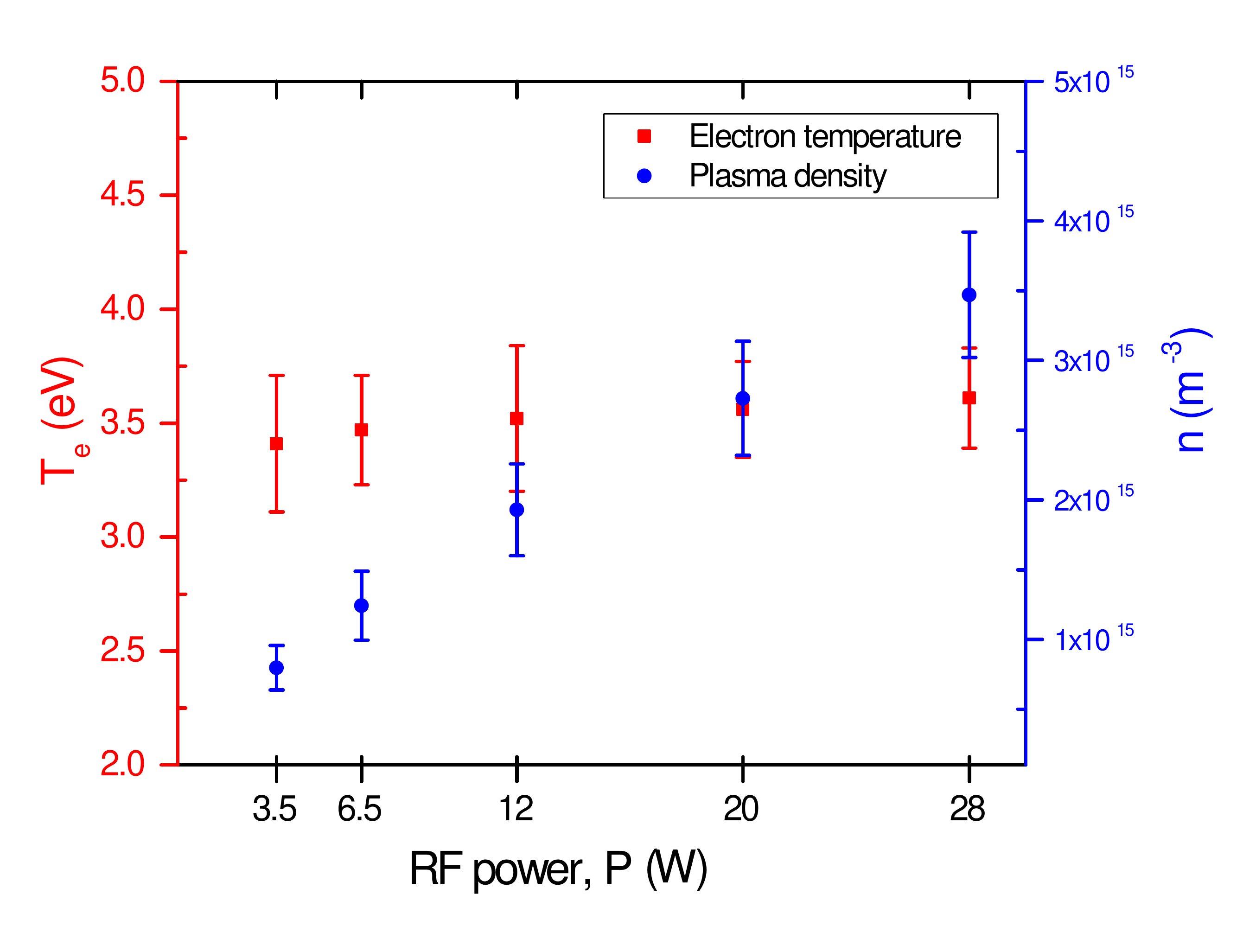}
\caption{\label{fig:fig4} Plasma density ($n$) and electron temperature ($T_e$) at different rf powers. The measurements are carried out at fixed argon pressure, p = 30 Pa in unmagnetized plasma.}
\end{figure}

Fig.~\ref{fig:fig2}(a) and Fig.~\ref{fig:fig2}(b) show the time-averaged $V_p$, $V_f$ and $V_s$ for different rf powers at constant pressure and for different pressures at constant power, respectively. We see a slight variation in the potentials at different discharge conditions in the absence of the magnetic field (B = 0 T). To see the effect of the object size on $V_s$ at a given discharge condition (P = 12 W and p = 30 Pa), the floating potential of the stainless steel spherical probes of different sizes are measured. The variation of $V_s$ for different sized spherical probes is depicted in Fig.~\ref{fig:fig3}. For getting the theoretical values of $V_s$ for the given discharge conditions, the plasma density ($n$) and electron temperature are measured using the double probe \cite{doubleprobemalter,doubleprobemagnetizedplasma,mangilalthesis}. 
The double probe is made of two tungsten wires (or single probes) of radius 0.15 mm and length 8 mm. Both probes are separated by $\sim$ 7 mm. At p = 30 Pa, the  plasma is moderately collisional; therefore the collisionless OML theory\cite{mottsmitomltheory1,chenprobe1} of cylindrical probe underestimates the plasma density. To measure the approximate plasma density, the collisional model for the ion current to the cylindrical probe is used \cite{trichyprobe2,kundraprobe1}.  The variation of $T_e$ and $n$ with different rf powers is depicted in Fig.~\ref{fig:fig4}. At this discharge condition, $\lambda_{De}$ is $\sim$ 0.3 mm, which gives $2 <\rho< 6$. The theoretically estimated values corresponding to this $\rho$ \cite{willisfloatingpotential2} are plotted with experimental data (Fig.~\ref{fig:fig3}) and found to be in good agreement within the error of $\sim$ 10 \% , which is expected due to the plasma collisionality. It confirms that the surface potential of a spherical object depends on its size in the a low temperature unmagnetized plasma ($T_e >> T_i$).
\subsection{Surface potential of spherical probes in magnetized plasma} 
\begin{figure*}
 \centering
\subfloat{{\includegraphics[scale=0.30]{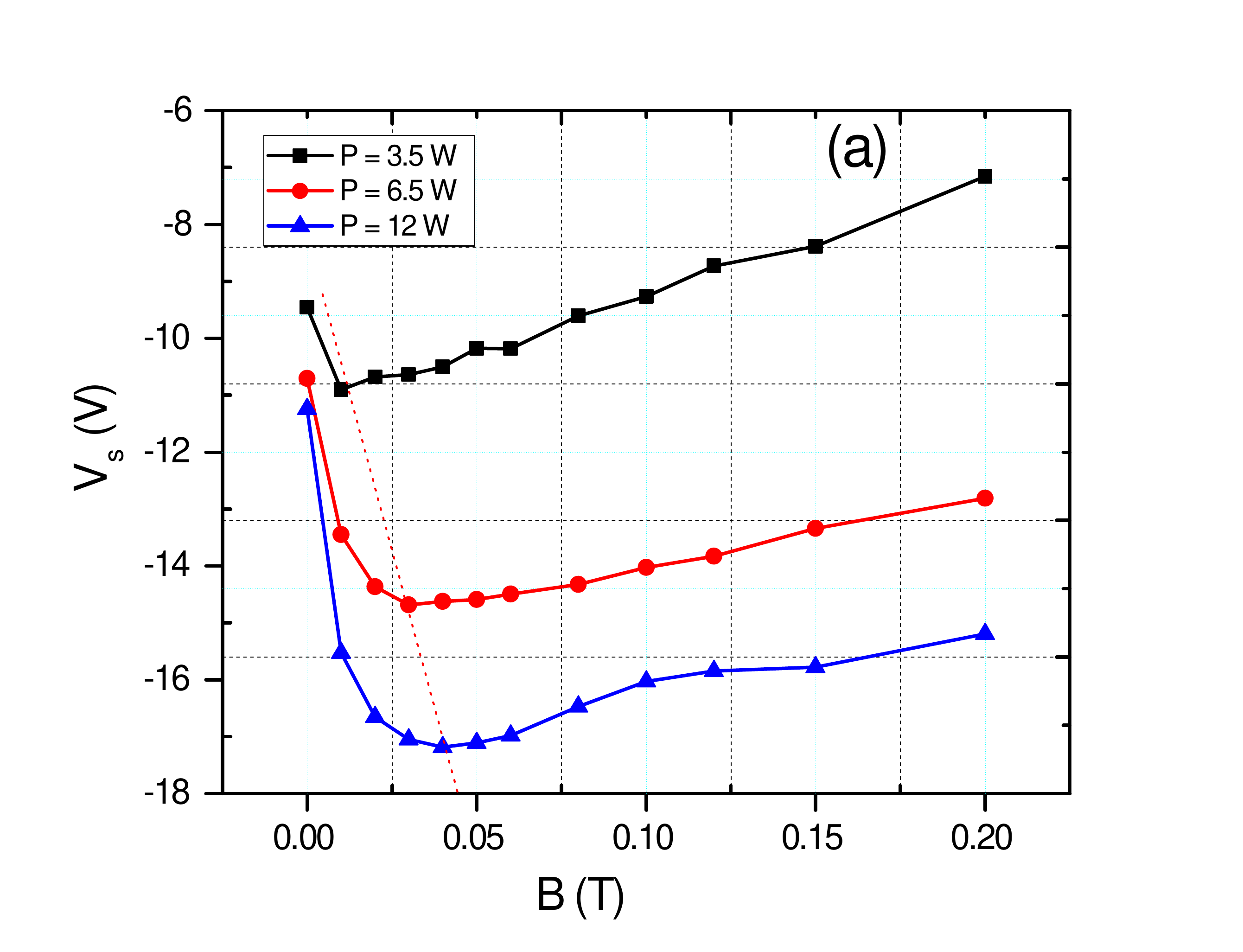}}}%
 \subfloat{{\includegraphics[scale=0.300]{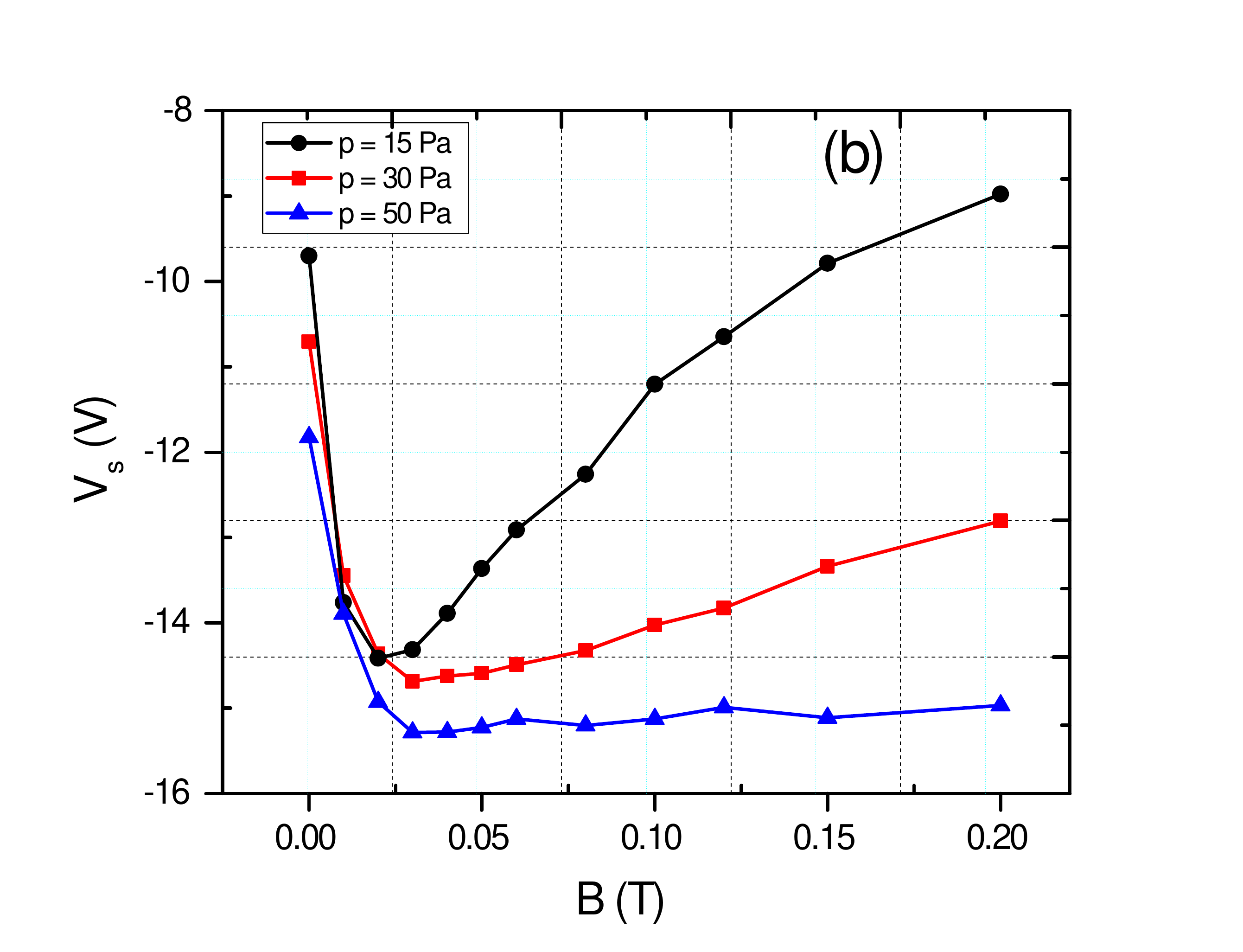}}}
 \qquad
\subfloat{{\includegraphics[scale=0.300]{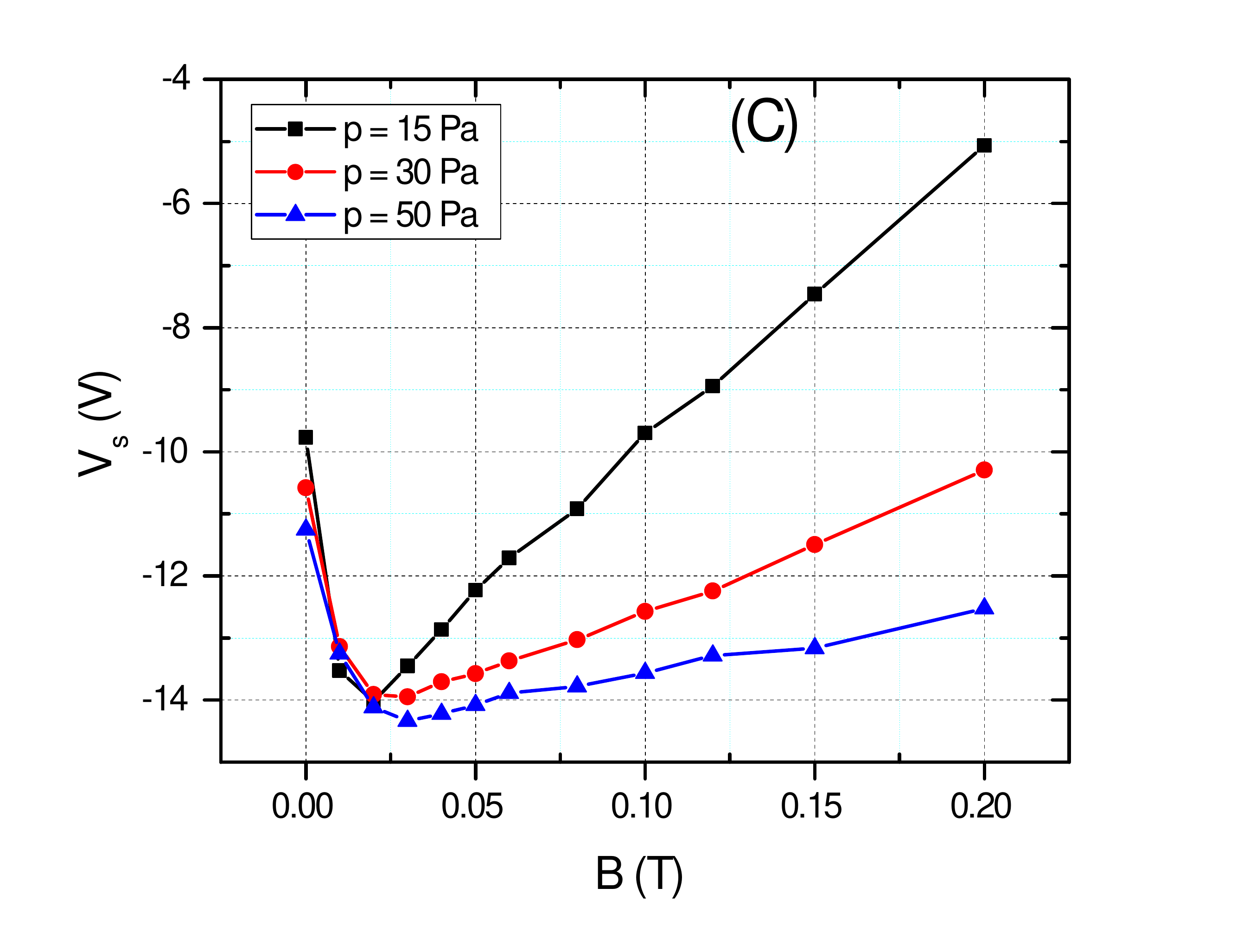}}}
 \qquad
 \caption{\label{fig:fig5}(a) Time--averaged surface potential ($V_s$) of the stainless steel spherical probe ($r$ = 1.25 mm) for different rf powers at fixed pressure, p = 30 Pa, in the plasma at different strengths of the magnetic field. The dotted line represents the shifting of the maxima of $V_s$ with increasing the input rf power. (b) Time-averaged $V_s$ of the stainless steel spherical probe ($r$ = 1.25 mm) for different argon pressures at rf power, P = 12 W in magnetized plasma. (c) Time--averaged $V_s$ of the bronze spherical probe ($r$ = 1.5 mm) for different argon pressures at rf power, P = 12 W in the plasma at different strengths of magnetic field. Here, the plotted values of $V_s$ are averaged over few data sets at given discharge condition.} 
 \end{figure*}

For producing the magnetized plasma, a B-field perpendicular to the plane of electrode (in Z-direction) is applied. In Fig.~\ref{fig:fig5} the surface potential of the SS probe of radius 1.25 mm and bronze probe of radius 1.5 mm at various strengths of the magnetic field are presented. It should be noted that the B-field is uniform in the entire plasma region at B = 0.2 T. The plots in Fig.~\ref{fig:fig5}(a) show the variation of $V_s$ for different input rf powers, P =  3.5, 6.5, and 12 W at a fixed pressure, p = 30 Pa. It is clearly seen in this figure that $V_s$ first increases (becomes more negative) at low B (B $<$ 0.05 T), attains a maximum value and after that it starts to decrease (becomes less negative) at higher magnetic field strength (B $>$ 0.05 T). The rate of change of $V_s$ is observed to be different in the low B-field region (B $<$ 0.05 T) and high B-field  region (B $>$ 0.05 T) at a given input power. It is also noticed that $V_s$ attains its maximum value at low B-field at lower input power (P = 3.5 W) and at high B-field at higher input power (P = 12 W).\par
%
The variation of $V_s$ at a given power (P = 12 W) and different pressures, p = 15, 30 and 50 Pa with the magnetic field strength is presented in Fig.~\ref{fig:fig5}(b). The rate of change of $V_s$ is less at higher pressures for B $>$ 0.05 T. It confirms the $V_s$ dependence on the plasma collisionality. Moreover, $V_s$ achieves its maximum value at lower B if the gas pressure is reduced. A similar trend of $V_s$ variation is observed for the different sized magnetic spherical probes in the presence of B-field. The variation of $V_s$ against B for a non-magnetic sphere (bronze) is depicted in Fig.~\ref{fig:fig5}(c). The surface potential shows a similar trend to that of the magnetic sphere (Fig.~\ref{fig:fig5}(b)) in the presence of a B-field. However, the rate of change of $V_s$ for the non-magnetic and magnetic sphere are different at the same discharge conditions.
 \subsection{Comparison of surface potentials}
\begin{figure*}
 \centering
\subfloat{{\includegraphics[scale=0.30]{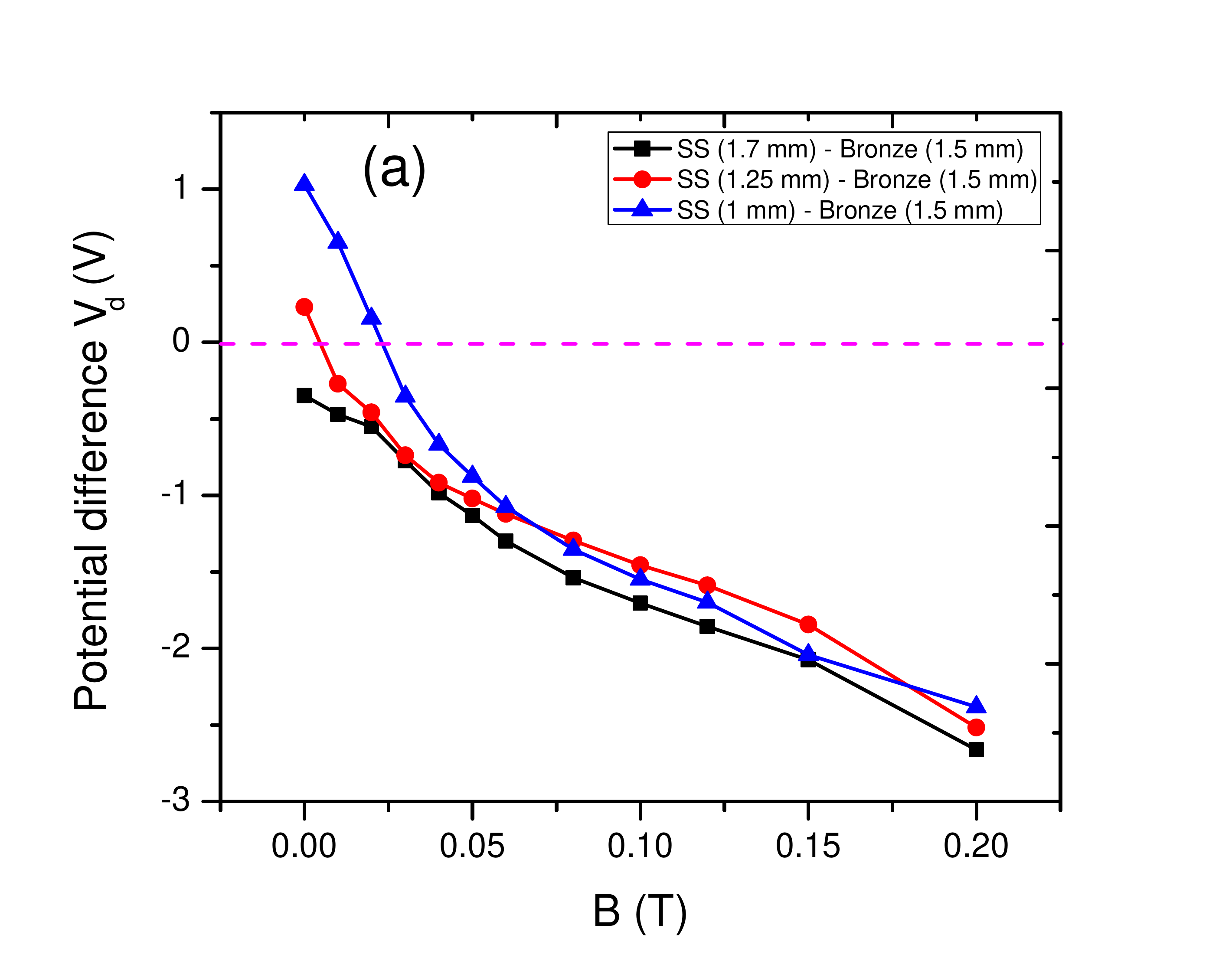}}}%
 \subfloat{{\includegraphics[scale=0.300]{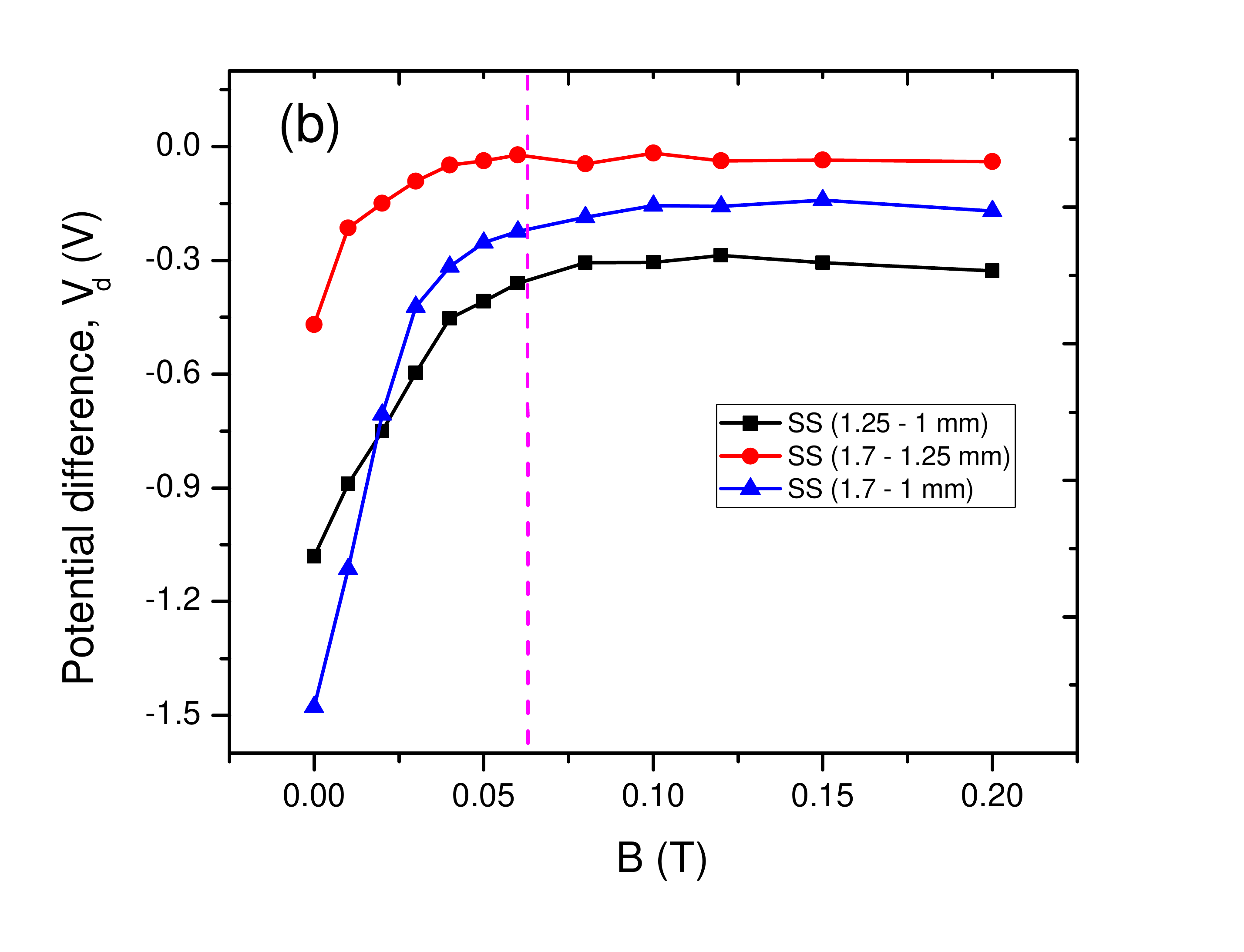}}}
 \qquad
 \caption{\label{fig:fig6}(a) The surface potential difference ($V_d$) for non-magnetic (bronze) and magnetic (SS) spherical probes at pressure, p = 30 Pa and rf power, p = 12 W in the plasma at various strengths of the magnetic field. (b) The difference of $V_s$ for different sized magnetic (SS) spherical probes at pressure, p = 30 Pa and rf power, P = 12 W in magnetized plasma.} 
 \end{figure*}

The $V_s$ data of the magnetic probes and the non-magnetic probe are compared to see the fundamental differences in the charging mechanism in a magnetized plasma. Comparison of $V_s$ for the magnetic and non-magnetic spheres at different B-field is depicted in Fig.~\ref{fig:fig6}. We take simultaneous measurements of the floating potential of a pair of spherical probes at a given discharge condition. At the same discharge condition, the floating potential is measured for different pairs of spherical probes with different sizes. It should be noted that the error in $V_f$ measurement is $< \pm$ 2 \% and the reference potential ($V_p$) is common for all measurements at similar discharge conditions. This gives almost an accurate trend of the potential difference. In Fig.~\ref{fig:fig6}(a) the surface potential of the bronze probe has been subtracted from the magnetic ones. It is reconstructed from the $V_s$ data for different sized ($r$ = 1.0 mm, 1.25 mm and 1.7 mm) SS and bronze ($r$ = 1.5 mm) probes to compare the size dependence in the presence of the B-field. It is clear from Fig.~\ref{fig:fig6}(a) that the smaller sized magnetic sphere (e.g., $r$ = 1 mm) has a higher value of $V_s$ than the non-magnetic sphere (e.g., $r$ = 1.5 mm) above B $>$ 0.03 T. This difference in $V_s$ increases with increasing B-field. It means that equally sized magnetic and non-magnetic spherical objects or dust grains have different charges in a magnetized rf discharge. Fig.~\ref{fig:fig6}(b) compares the surface potential of the different sized magnetic spheres to see the role of the B-field in determination of $V_s$. It is seen in Fig.~\ref{fig:fig6}(b) that the difference in $V_s$ for different sized magnetic probes decreases with increasing magnetic field and remains almost constant at higher B-field (B $>$ 0.05 T). It shows that $\alpha$ remains almost constant for all sized spherical probes at higher B-field, B $>$ 0.05 T. In other words, $V_s$ has a much weaker size dependence in a magnetized rf plasma. 
 \section{Discussion} \label{sec:results_discussion}
 The surface potential of a spherical probe or dust grain is determined by the electron and ion currents to its surface. In a low temperature plasma, where $T_i \ll T_e$, the surface potential is mainly determined by $T_e$. Since $v_{the}\gg v_{thi}$, the surface potential is always negative with respect to the plasma potential. Here, $v_{the}$ and $v_{thi}$ are the electron and ion thermal velocities, respectively. In an unmagnetized rf discharge plasma (at B = 0 T), the surface potential of a spherical object of the radius $r > \lambda_{De}$ is estimated using the theoretical value of $\alpha$ in the transition region between OML and TS regions \cite{willisfloatingpotential, willisfloatingpotential2}. A slight variation in $T_e$ (see Fig.~\ref{fig:fig4}) with increasing the power and pressure demonstrates a negligible change in $V_s$ in the unmagnetized plasma. (see Fig.~\ref{fig:fig2}).\par
 With the application of a magnetic field, the gyro-radius of electrons ($r_{ge} = m_e v_{the}/e B$) and of ions ($r_{gi} = m_i v_{thi}/e B$) decreases with increasing B-field. Due to the mass differences, electrons are magnetized at lower magnetic field than ions, i.e., $r_{ge} \ll r_{gi}$. The electrons and ions are considered to be magnetized when the gyration frequency of the respective species ($\omega_{ce/ci}$) is higher than the collisonal frequency ($\nu_{e-n/i-n}$), i.e., $r_{ge/i}<\lambda_{e-n/i-n}$. Here $\lambda_{e/i}$ is the collisional mean free path for the respective species. In our experiments (p = 15 to 50 Pa and P = 3.5 to 12 W), $n$ and $T_e$ are observed to be vary between $\sim$ 6$\times 10^{14}~m^{-3}$ to 3$\times 10^{15}~ m^{-3}$ and 3--5 eV, respectively. The mean free path of electrons, $\lambda_{e-n} = 1/n_g \sigma_{e-n} \sim$14-3~mm and ions, $\lambda_{i-n} = 1/n_g \sigma_{i-n} \sim$ 0.2-0.08~mm. Here $\sigma_{e-n} \sim 2\times 10^{-20}~ m^{2}$ and $\sigma_{i-n} \sim 1\times 10^{-18}~ m^{2}$ are the collision cross sections of electrons and ions with argon atoms, respectively\cite{collisionacrosssection} and $n_g$ is the neutral gas density. The electron gyro-radius varies between $r_{ge} \sim$ 0.5-0.7~mm for B = 0.01 T at given discharge conditions. Therefore, the condition $r_{ge} < \lambda_{e-n}$ meets even below B = 0.01 T. With increasing the strength of the magnetic field (B $>$ 0.01 T), $r_{ge}$ continuously decreases and electrons are fully magnetized. Ions are assumed to be at room temperature, i.e, $T_i \sim 0.03 ~eV$. The ion gyro-radius $r_{gi}$ at B = 0.2 T is estimated as $\sim$ 0.5 mm, which indicates that ions get magnetized at high magnetic field, B $>$ 0.2 T. It essentially means that in the range of the magnetic field (B $<$ 0.2 T), only electrons are magnetized but ions are assumed to be unmagnetized. In a magnetized plasma, the currents $I_e$ and $I_i$ to the surface of a spherical probe are altered when the condition, $r_{ge/i} < \lambda_{De}$, is satisfied. Here $\lambda_{De}= \sqrt{\epsilon_0 k_B T_e/e^2 n_e}$ is the electron Debye length. In the present work, $\lambda_{De}$ varies between $\sim$ 0.2 to 0.5 mm for the given range of plasma parameters. It shows that the electron current gets changed as a B-field is introduced. The ions do not fulfil this criteria at B $<$ 0.2 T, therefore, the ion current to the surface of the spherical object is considered to be unaffected. \par
\begin{figure*}
 \centering
\subfloat{{\includegraphics[scale=0.32]{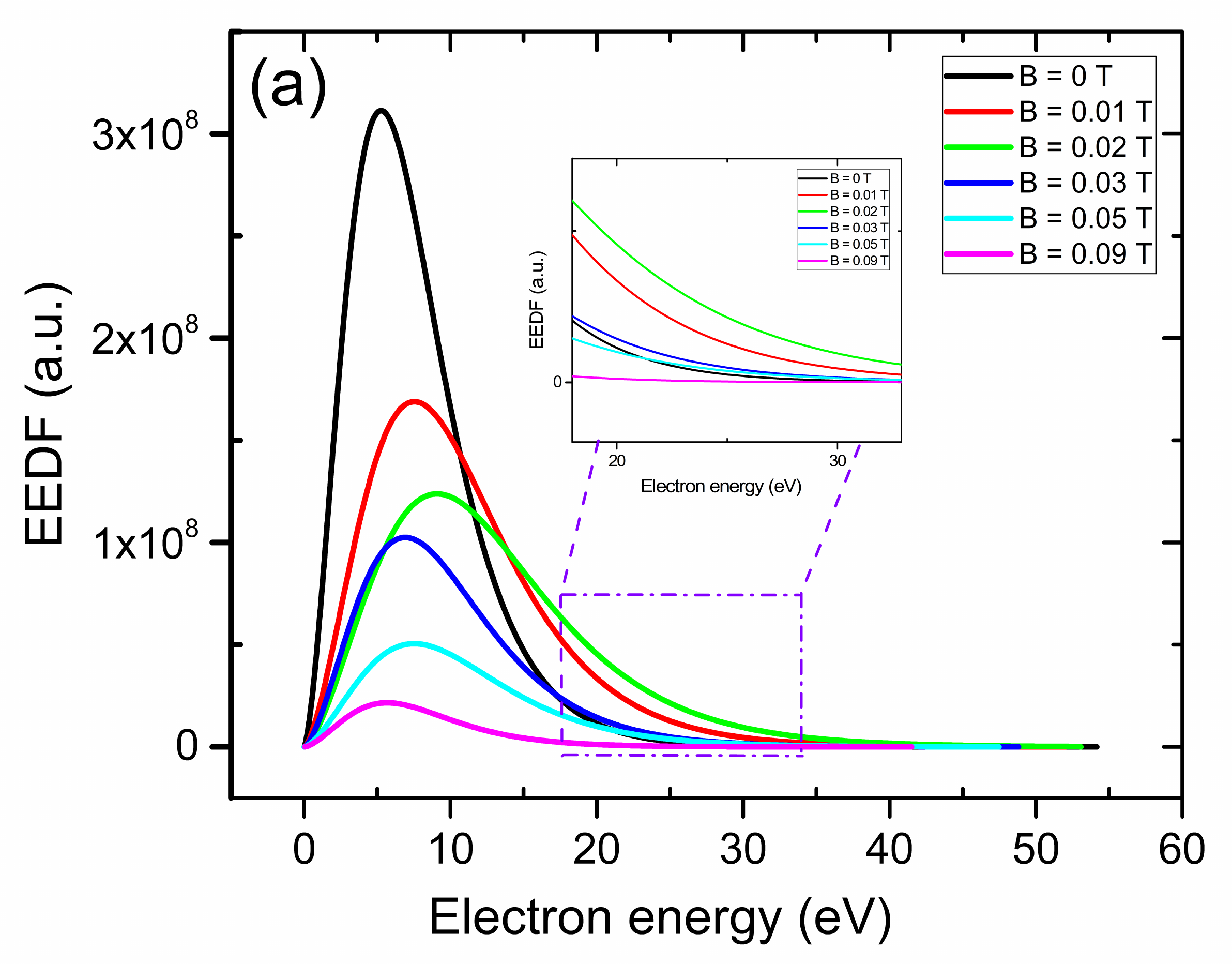}}}%
 \subfloat{{\includegraphics[scale=0.32]{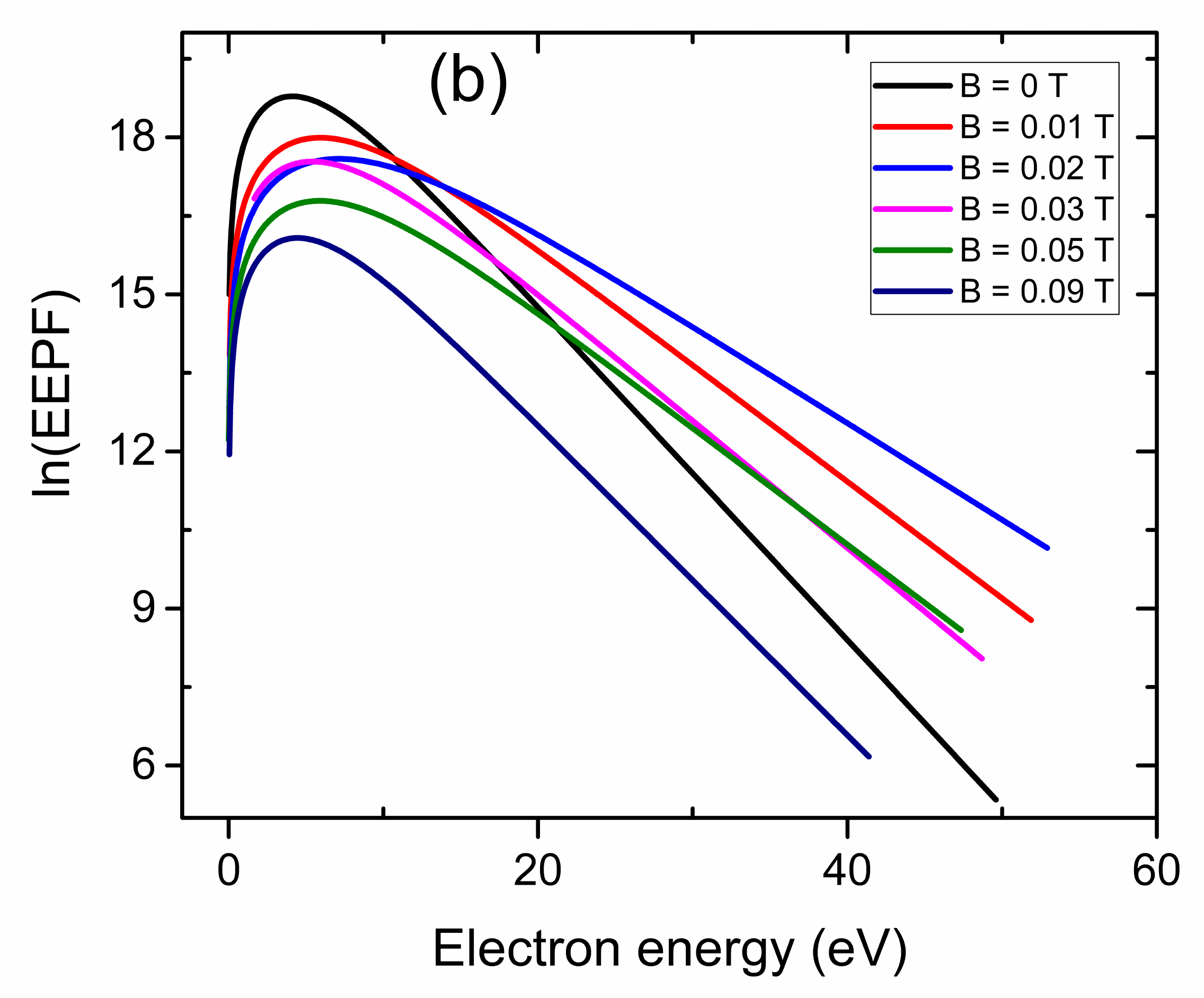}}}
 \qquad
 \caption{\label{fig:fig7}(a) Electron energy distribution function (EEDF) with external magnetic field at gas pressure, p = 30 Pa and rf power, P = 12 W. The inset image represents the enhancement of the energetic electron population with the application of a magnetic field. (b) The plots of ln(EEPF) for given EEDF at various strengths of the magnetic field.} 
 \end{figure*}
\begin{figure}
 \centering
\includegraphics[scale=0.355001]{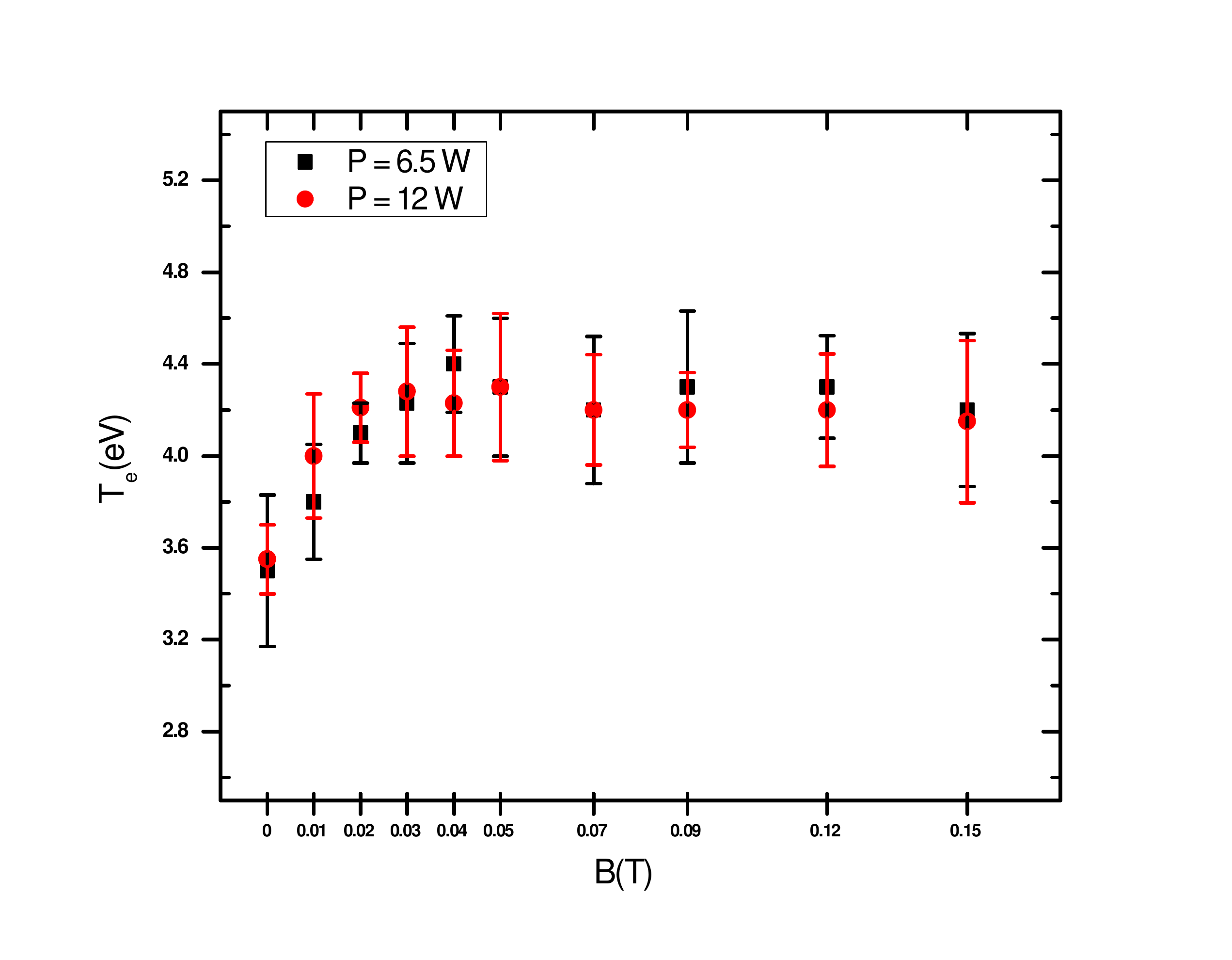}%
 \caption{\label{fig:fig8}Electron temperature ($T_e$) variation with magnetic field at pressure, p = 30 Pa and rf powers, P = 6.5 and 12 W.} 
 \end{figure}
In an unmagnetized plasma, a constant flux of energetic electrons is lost to the chamber wall. The magnetic field confines the electrons, which definitely reduces the electrons loss to the chamber wall. Therefore, it is expected that the density of the energetic electrons would be increased as the magnetic field is introduced. To see the effect of the B-field on the electron population, the electron energy distribution function (EEDF) is measured using a tungsten cylindrical probe of length $l_p =$ 8 mm and radius $r_p =$ 0.15 mm. The probe is positioned perpendicular to the discharge axis or magnetic field lines. At discharge condition (P = 12 W and p = 30 Pa), $r_p <\lambda_{De}$ therefore, the conventional probe theory of a cylindrical probe is used to get the EEDF. It should be noted that the plasma anisotropy in presence of the magnetic field depends on the parameter B/p and this value should be higher than $3\times 10^{-2}$ T/Pa \cite{Behnkeanisotropyeedf}. Since in our set of experiments, the ratio of B/p varies from $\sim$ $1\times 10^{-4}$ to $2\times 10^{-3}$ T/Pa, it does not exceed $3\times 10^{-2}$ T/Pa. Therefore, we do not expect any substantial anisotropy of the plasma or EEDF in our measurements. It should also be noted that the second derivative probe method gives a reliable EEDF in the range of the diffusion parameter \cite{eedfdiffusion} $\Psi = r_p \frac{ln(\pi l_p/4 r_p)}{\gamma r_{ge}}< 30$ \cite{eedfsecondderivativeinb}. Here, $\gamma$ is constant and we assumed $\gamma$ $\sim$ 4/3 for our pressure regime. In our case, diffusion parameter has the value $\Psi$ $<$ 15 for B $<$ 0.1 T; therefore, this method is used to get the EEDF to show the increase in the energetic electron population as the B-field is turned on. The EEDF is estimated from the second derivative of the probe I-V characteristics with respect to the probe voltage \cite{godyakeedfbook,kundraprobe1, eedfsecondderivativeinb}, 
\begin{equation}
 F(E)= \frac{2 \sqrt{2 m_e}}{A_p e^3} \sqrt{E}  \dfrac{d^2I_e}{dV^2}
 \end{equation} 
where $E$ = eV = e($V_p - V_b$), $I_e$ is the electron current to probe, $V_b$ is the probe bias, $V_p$ is the  plasma potential and $A_p$ is the area of probe. 
 The EEDF with magnetic field at p = 30 Pa and P = 12 W is shown in Fig.~\ref{fig:fig7}(a). The population of the cold (or lower energy) electrons, which are reaching the probe, decreases with increasing B-field whereas the population of the energetic electrons increases at low B $<$ 0.05 T (see inset image). It means that the energetic electrons can easily reach this probe surface at low B-field.\par
It is obvious that the average electron energy will increase due to the increase of the density of energetic electrons. It means $T_e$ is expected to increase while the magnetic field is introduced. Since the variation in $T_e$ affects the surface potential of a spherical object (see equation 3), it is measured using a double or single probe at various strengths of the B-field. The inverse slope of $ln(EEPF)$ = $ln(\frac{F(E)}{\sqrt{E}})$ with respect to $E$ (see Fig.~\ref{fig:fig7}(b)) gives $T_e$. It should be noted that the single probe used to obtain the EEDF is not compensated and overestimates $T_e$, therefore, errors concerning $T_e$ is expected at low B-field. At higher B (B $>$ 0.09 T), the secondary plasma around the probe tip during the positive bias does not give true I--V characteristics, which is also a cause error in $T_e$ measurement even though it is rf compensated. In view of this, a double probe is used to obtain the approximate value of $T_e$ upto B $\sim$ 0.15 T. The double probe theory \cite{doubleprobemalter} estimates reliable plasma parameters ($n$ and $T_e$) in rf discharges if electrons obey the Maxwellian distribution, i.e., the EEDF should be Maxwellian in the presence of the B-field. In Fig.~\ref{fig:fig7}(b), $ln(EEPF)$ is plotted against the electron energy E for different values of B. The $ln(EEPF)$ against E shows the characteristics of a Maxwellian plasma \cite{godyaknonmaxwellianplasma} in the experiments. Fig.~\ref{fig:fig8} represents the variation of the electron temperature ($T_e$) with magnetic field at p = 30 Pa and P = 6.5 and 12 W. \par
In the magnetized plasma, the net electron current $I_e$ to the spherical probe is a sum of the electron currents to different positions on the probe \cite{dustcurrent}. In the present experiments, it is difficult to estimate the electron current as a function of position (with respect to the magnetic field direction) on the spherical probe surface \cite{dustcurrent}.
Therefore, a simple model by considering the net electron current ($I_e$) in two possible directions along the B-field ($I_{e\parallel}$) and transverse to the B-field ($I_{e\perp}$) is used to explain the observed results qualitatively. Here $I_{e\parallel}$ and $I_{e\perp}$ are assumed to be the electron current components at a given position on the spherical probe with respect to the magnetic field direction. The total electron current to the surface of a spherical probe is $I_e$ = $I_{e\parallel}$ + $I_{e\perp}$, which determines the floating surface potential of a probe in a magnetized plasma. It should be noted that the electron motion transverse to the B-field is much more hindered than that of along the B-field in the moderately collisional plasma. In other words, $I_{e\perp}$ is reduced much more than $I_{e\parallel}$ in a magnetized plasma \cite{bohmprobeinmagneticfield,probeinstrongb}, i.e., $D_{e\perp} < D_{e\parallel}$, where $D_{e\perp}$ and $D_{e\parallel}$ are the transverse and longitudinal diffusion coefficients, respectively. \par
\begin{figure}
\centering
 \includegraphics[scale= 0.34000]{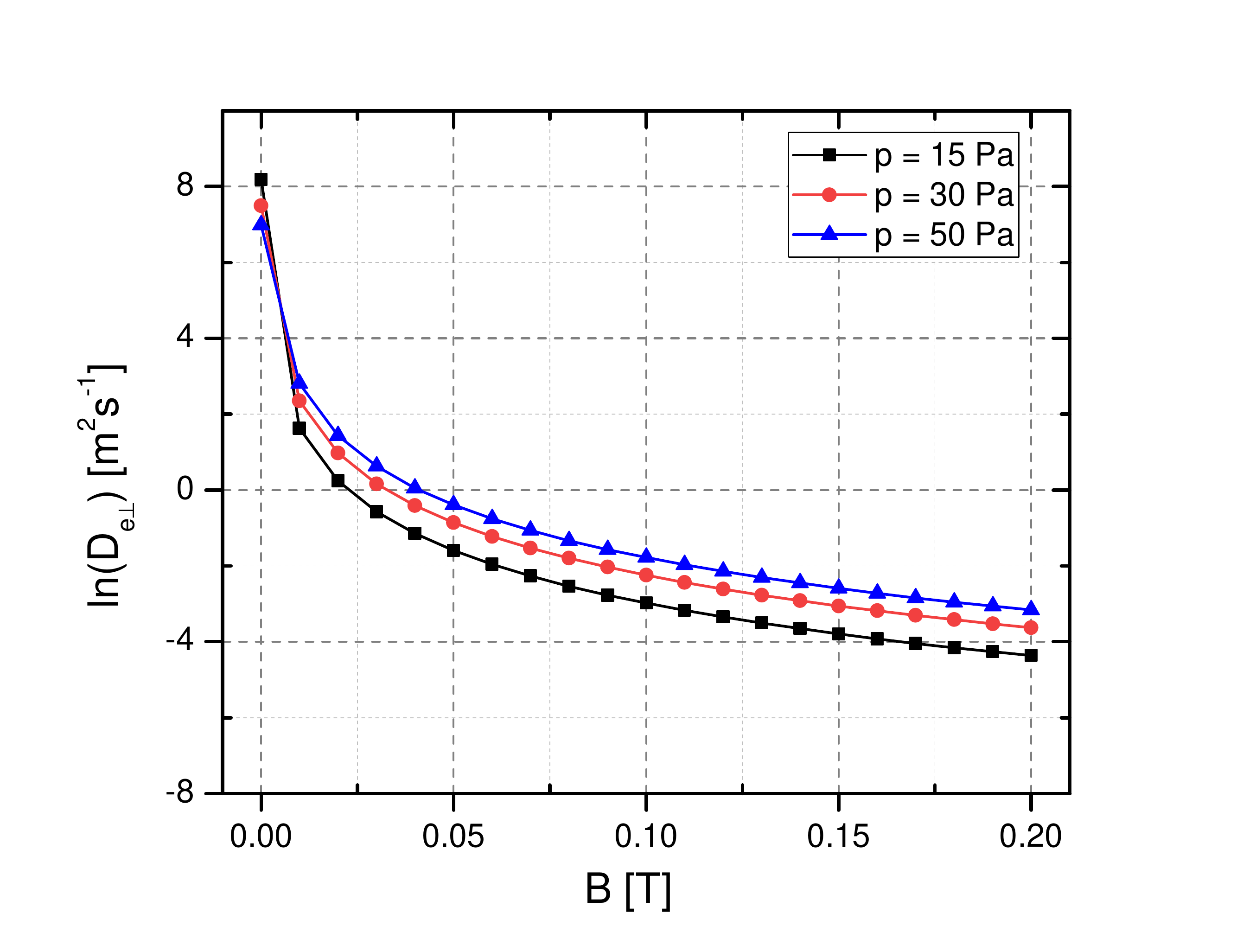}
\caption{\label{fig:fig9} Logarithm plots of transverse diffusion coefficient ($D_{e\perp}$) for different argon pressures at various strengths of magnetic field.}  
\end{figure}

There are two possible diffusion processes: the first one is the drain diffusion and the second one is the collisional diffusion. In drain diffusion, electrons may change their direction and cross the B-field during the motion in rf oscillating sheath field of a spherical object. In collisional diffusion, the gyrating electrons collide with background neutrals and diffuse across B-field with a higher rate \cite{bohmprobeinmagneticfield}. The drain diffusion mainly dominates over the collision diffusion in a low pressure magnetized plasma. Since the present work is performed in a moderately collisional plasma, the collisional diffusion process is considered to be more effective. For moderately collisional low temperature magnetized plasmas, the collisional transverse diffusion coefficient is $D_{e\perp} = D_{e0}/(1 + \omega^2_{ce} \tau^2_e)$, where $D_{e0} = \lambda_{e-n} v_{the}/3$ is the diffusion coefficient in the absence of a B-field, $\omega_{ce} = e B/m_e$ is the electron cyclotron frequency and $\tau_e = \lambda_{e-n}/v_{the}$  is the electron--neutral collision time \cite{chendiffusioncoefficient,chenplasmaphysicsbook,bohmprobeinmagneticfield}. \\
In a collisionless magnetized plasma, electrons motion is restricted by the magnetic field in the transverse direction. Therefore, the $D_{e\parallel}$(or $I_{e\parallel}$) play a dominant role in determination of the surface potential (or charge) of the dust grain in the presence of magnetic field \cite{dustcurrent}. The logarithmic plots of transverse diffusion coefficient, $ln({D_{e\perp})}$, is shown in Fig.~\ref{fig:fig9} which show a reduction in $D_{e\perp}$ after a magnetic field strength of 0.01 T. The value of $D_{e\perp}$ increases with the pressure while the magnetic field strength (B $>$ 0.02 T) is kept constant, which is also illustrated in Fig 9.\par
In the low magnetic field regime (B $<$ 0.05 T), the increase in $T_e$ (see Fig.~\ref{fig:fig8}) enhances the surface potential (more negative) of a spherical probe according to equation 3. Now, the probe collects a higher net current ($I_e$) than $I_{e0}$. Here, $I_{e0}$ represents an equilibrium electron current to the spherical probe in unmagnetized plasma. This can also be understood on the basis of energetic electron population.    
Experimentally, the dominating role of energetic electrons in the charging process of a spherical object or dust grain in the plasma has been confirmed\cite{arnasductcharging}. Since the Larmor radius of the energetic electrons ($T_e >$ 20 eV) lies between 1.5 mm to 0.4 mm for B $<$ 0.05 T, energetic electrons are considered to be weakly magnetized. Since $D_{e\perp}$ decreases in this B-field regime (B $<$ 0.05 T), the net electron current to spherical surface should be lowered. However, an opposite behaviour (large $I_e$ or $V_s$) is observed at low B-field because of the confinement of energetic electrons (or higher $T_e$). The higher charges on the dust grains or more negative surface potential in a weakly magnetized plasma is also observed numerically by Tomita \textit{et al.}\cite{tomitadustchargingwithmagneticfield}. They claimed a larger absorption cross section for electron capture on the dust surface in the presence of a magnetic field. In the present study, higher charges on a spherical probe is due to the increase of $T_e$ for the higher magnetic field (see Fig.~\ref{fig:fig8}). \par
\begin{figure}
\centering
\subfloat{{\includegraphics[scale=0.33]{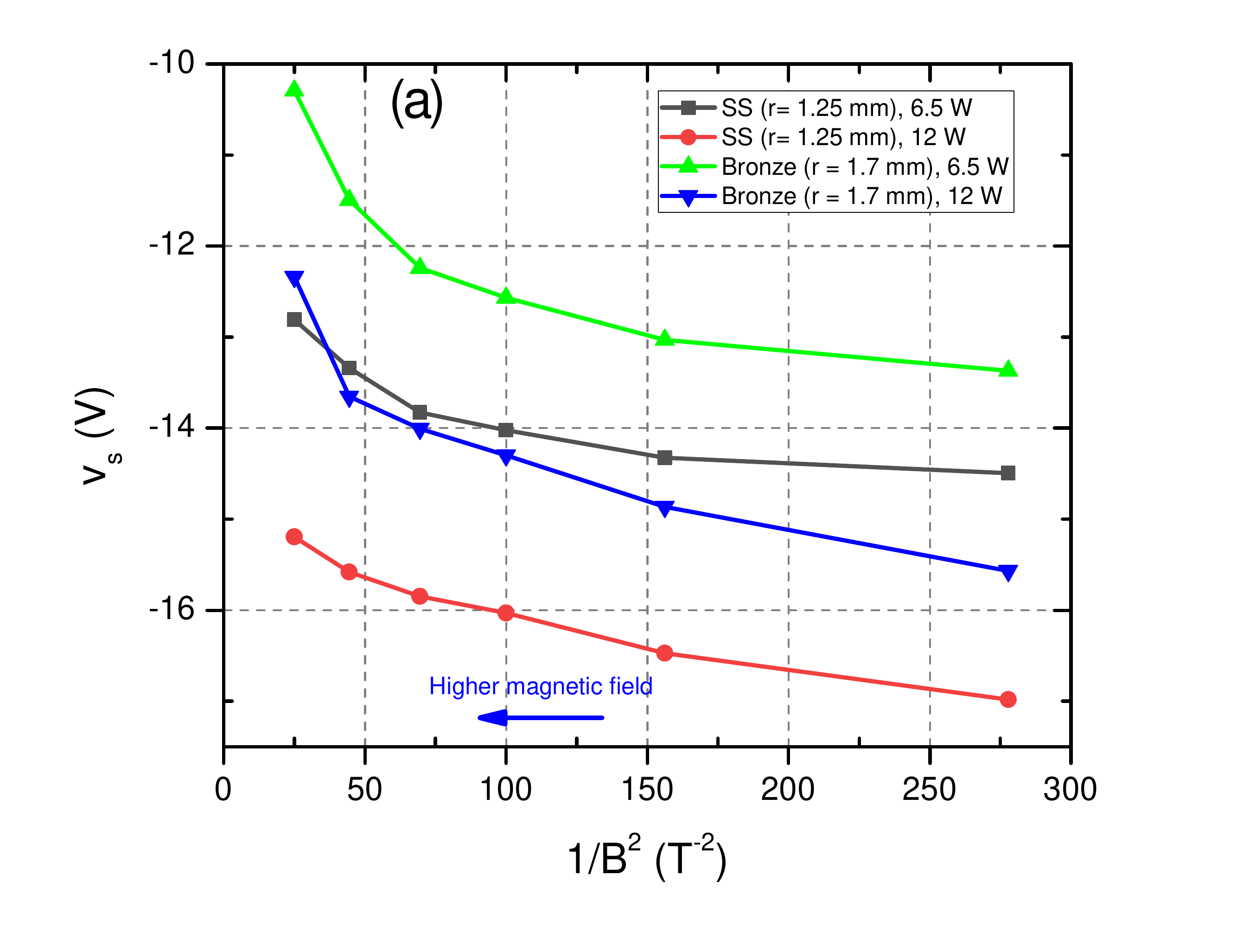}}}%
 \qquad
 \subfloat{{\includegraphics[scale=0.33]{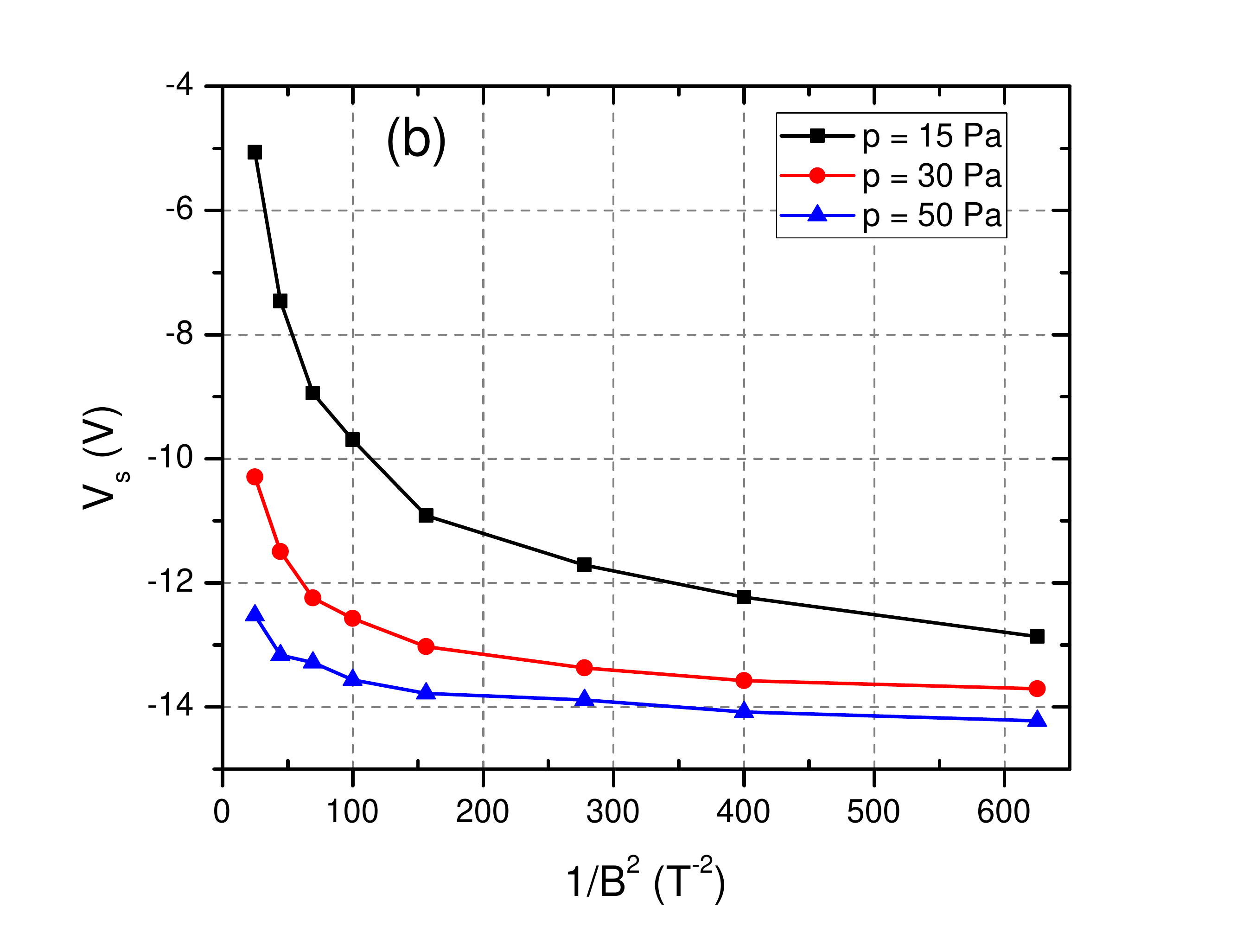}}}
\caption{\label{fig:fig10}(a) Surface potential variation with $1/B^2$ for different spherical probes at pressure p = 30 Pa and rf powers, P = 6.5 W and 12 W. The values of $1/B^2$ are correspond to B = 0.06 T to 0.2 T (b) $V_s$ variation against $1/B^2$ for bronze probe ($r$ = 1.7 mm) at rf power 12 W and different pressures. The values of $1/B^2$ are correspond to B = 0.04 T to 0.2 T}.  
\end{figure}
At higher B-field (B $>$ 0.05 T), a lower value of the mean free path ($r_{ge} < \lambda_{e-n}$) increases the electron-neutral collision frequency, resulting in a reduction of $T_e$. A slight reduction in $T_e$ at higher B is seen in Fig.~\ref{fig:fig8}. At fixed gas pressure, transverse diffusion coefficient ($D_{e\perp}$) decreases with the increasing magnetic field (see Fig.~\ref{fig:fig9}) which causes reduction in $I_{e\perp}$. It is also observed in numerical simulations that $I_{e\parallel}$ start to decrease with increasing the B-field. However, this reduction in $I_{e\parallel}$ is small compared to the reduction in $I_{e\perp}$ in the magnetic field regime (B $<$ 0.15 T). Therefore, it is assumed that the reduction in $I_{e\perp}$ to the spherical probe could be a possible cause of lower or less negative surface potential at high B-field (B $>$ 0.05 T), which is shown in Fig.\ref{fig:fig5}. \par
Since $D_{e\perp}$ (or $I_{e\perp}$) varies with $1/B^2$ at a given pressure, $V_s$ should has a $1/B^2$ dependence. In Fig.~\ref{fig:fig10}(a), $V_s$ is plotted against $1/B^2$ for different spherical probes between B = 0.06 T to 0.2 T. Fig.~\ref{fig:fig10}(a) clearly indicates that $V_s$ decreases linearly with $1/B^2$ between B $\sim$ 0.06 T to 0.15 T for an rf power of 12 W. However, the upper B-field value shifts to a slightly lower value (B $>$ 0.10 T) for the low power discharge (P = 6.5). Since plasma density is different in both cases (P = 12 W and 6.5 W) at p = 30 Pa, different rates of reduction of $V_s$ with $1/B^2$ are expected. The plots of $V_s$ against $1/B^2$ for different pressures at given power (P = 12 W) is shown in Fig.~\ref{fig:fig10}(b). We also see a linear variation of $V_s$ with $1/B^2$  below B $<$ 0.12 T and a non-linear reduction in $V_s$ above B $>$ 0.12 T (above p $\geq$ 30 Pa). For low pressure case (p = 15 Pa), the non-liner behaviour of $V_s$ is observed at magnetic field of $>$ 0.10 T. The different linear rates of $V_s$ at different pressures (Fig.~\ref{fig:fig10}(b)) are expected because of different plasma backgrounds (plasma density and $T_e$) at a fixed input rf power and different gas pressures. In a similar plasma background, we could expect a constant linear rate of $V_s$ with $1/B^2$ at different pressures according to the theoretical estimation as shown in Fig.~\ref{fig:fig9}. It shows that the reduction in $I_e$ (or $V_s$) is mainly due to the lower value of $I_{e\perp}$ in this magnetic field regime.\par
At strong magnetic field strength, the contribution of $I_{e\perp}$ to $I_e$ starts to reduce and a larger contribution comes from $I_{e\parallel}$ \cite{probeinstrongb}. Hence the role of $D_{e\parallel}$ (or $I_{e\parallel}$) becomes more important in determination of the surface potential. In the present experiments, $V_s$ does not decrease linearly with $1/B^2$ above B $>$ 0.12 T (Fig.~\ref{fig:fig10}) except for the low density plasmas (p = 30 Pa, P = 6.5 W and p = 15 Pa, P = 12 W). In moderately collisional plasma, a reduction in $I_{e\parallel}$ is expected while the magnetic field increases but it could be comparable or larger than of $I_{e\perp}$ at strong B-field  \cite{probeinstrongb}. The non-linear characteristics of $V_s$ against $1/B^2$ above B $>$ 0.12 T (see  Fig.~\ref{fig:fig10}) is expected due to a dominant role of $I_{e\parallel}$ along with $I_{e\perp}$ for determining the surface potential. In some other experiments and simulations a reduction in $V_s$ of a spherical object with increasing magnetic field has also been reported.
\cite{melzermagnetizeddusty,langefloatinginmagnetized,dotecharginginmagneticfield,particlechargemarkus1}.\par
It is known that the current $I_{e\perp}$ to the spherical probe decreases with increasing $\omega_{ce} \tau_e$ or decreasing $D_{e\perp}$. At given finite B-field, $D_{e\perp}$ increases with increasing the electron-neutral collisions or gas pressure, which leads to an increase of $I_{e\perp}$. At finite B-field (B $>$ 0.01 T), $D_{e\perp}$ has slightly larger value at higher pressure (p = 50 Pa) than at lower pressure (p = 15 Pa), as shown in Fig.~\ref{fig:fig9}. The difference in $D_{e\perp}$ for different pressures at finite B-field is one of the possible causes for the different values of $V_s$ in the presence of a magnetic field (Fig.~\ref{fig:fig5}(b)). The spherical surface collects more electron current at higher pressure (p = 50 Pa) because of the large value of $D_{e\perp}$. The lower value of $D_{e\perp}$ causes a smaller electron current to the spherical probe. Therefore, the spherical probe has a higher value (more negative) of $V_s$ at p = 50 Pa than at p = 15 Pa (Fig.~\ref{fig:fig5}) in the magnetized plasma (B $>$ 0.05 T).  These results confirm the dominating role of the collisional diffusion over the drain diffusion in a moderately collisional magnetized plasma.\par
 It should be noted that in an unmagnetized plasma (B = 0 T), the dust charge or surface potential decreases with increasing the pressure due to the higher ion-neutral collision frequency \cite{kharpakdustneutralcharge}. It is also known that collisions of plasma species with neutrals retard the motion along the magnetic field, hence reduction in $I_{e/i \parallel}$ is expected with increasing the gas pressure. Since ions are unmagnetized below the magnetic field of $<$ 0.2 T, the role of ions is negligible than the electrons in determination of the surface potential in this given magnetic field regime (B $<$ 0.2 T). Due to the smaller gyro-radius of electrons (higher gyro-frequency) above B $>$ 0.05 T, the electron-neutral collision frequency is found to be larger. It means that collisions lower the $I_{e\parallel}$ more effectively than $I_{i \parallel}$ in the magnetized discharge. Hence the role of ion current for determining the surface potential at different pressures (p = 15 to 50 Pa) is assumed to be negligible above B $>$ 0.05 T. The higher negative value of $V_s$ at higher pressure also confirms a dominant role of $I_{e\perp}$ in reduction of the net electron current ($I_e$) or the surface potential in the magnetized plasma. \par 
The difference in $V_s$ for magnetic (stainless steel) and non-magnetic (bronze) spherical objects (Fig.~\ref{fig:fig6}) is understood on the basis of field line distribution around a spherical body in a magnetized plasma. Since $I_{e\perp}$ decreases faster than $I_{e\parallel}$ with the magnetic field, $I_{e\perp}$ mainly responsible for the reduction in $I_e$ (or $V_s$). The magnetic flux density on either side of the magnetic sphere is less than that inside of it in the presence of a magnetic field (see Fig. 3.7 of ref \cite{magneticfieldlines}), which enhances $I_{e\perp}$ to the object surface due to the large value of $D_{e\perp}$ that varies with $1/B^2$. Thus, the electron current, $I_e$, to the magnetic sphere increases making the surface of the spherical object more negative (higher $V_s$) in the presence of B-field.
In the case of the non-magnetic object (copper bronze), the magnetic flux density on either side of the sphere is slightly larger than that inside of it due to the diamagnetic characteristics of copper. The B-field line density on either side of the non-magnetic sphere is expected to be higher than that of around the magnetic sphere, which reduces $I_{e\perp}$ to the object surface. The lower value of the electron current ($I_e$) to the non-magnetic sphere makes surface less negative. Hence a magnetic sphere has a higher value (more negative) of $V_s$ than that of a non-magnetic sphere (see Fig.~\ref{fig:fig6}(a)) above a finite value of B-field (B $>$ 0.03 T). \par
 The qualitative description presented here provides a better understanding of the observed surface potential variation for magnetic and non-magnetic spherical probes (or large dust grains) in a magnetized rf discharge. We have provided a possible charging mechanism of magnetic and non-magnetic particles based on the reduction of electron current to spherical surface. To the best of our knowledge, at present there is not an analytical or simulation work for a moderately collisional magnetized plasma to support our claim. Therefore it may be possible that another charging mechanism of magnetic and non-magnetic spherical particles plays a role in a magnetized plasma.  
 \section{Conclusion}  \label{sec:summary}
 The surface potential of magnetic (stainless steel) and non-magnetic (bronze) spherical objects in a magnetized rf discharge at various discharge conditions is measured. A 13.56 MHz rf generator is used to produce the plasma between a transparent indium tin oxide (ITO) coated glass electrode and a metal electrode. A superconducting electromagnet with Helmholtz coils configuration is used to introduce an external magnetic field. The vacuum chamber is placed at the center of the magnet to perform the experiments in a uniform magnetic field. The surface potential of different sized magnetic spherical probes ($r$ = 1.0 mm , 1.25 mm and 1.7 mm) is measured and compared with a non-magnetic spherical probe ($r$ = 1.5 mm) in the plasma at different strengths of the B-field. The main findings of the experimental studies are listed below: 
 \begin{enumerate}
 \item The surface potential ($V_s$) of the spherical object ($r >\lambda_{De}$) depends on its size in the unmagnetized plasma.
 \item The surface potential of a spherical object either magnetic or non-magnetic increases at the low magnetic field (B $<$ 0.05 T), attains a maximum value and starts to decrease with further increasing the strength of the external magnetic field (B $>$ 0.05 T). The rate of change of the surface potential in the magnetized plasma strongly depends on the gas pressure as well as the plasma parameters ($n$ and $T_e$).
 \item The surface potential of magnetic spherical objects or large dust grains is found to be higher (more negative) than that of a non-magnetic sphere at the higher magnetic field (B $>$ 0.04 T).
  \item The surface potential of the spherical objects loses its size dependence characteristics in the rf discharge with the application of an external magnetic field (B $>$ 0.05 T).
 \end{enumerate}
 The magnetic field reduces the loss of energetic electrons to the wall and confines them in the plasma volume. The average energy of bulk plasma electrons or $T_e$ increases due to the confinement of energetic electrons. An increase in energetic electron population or $T_e$ at lower B (B $<$ 0.05) increases the electron current to the probe surface. Hence, the surface potential increases (more negative) with B and attains its peak value between B = 0.01 T and 0.05 T for different discharge conditions. With increasing the B-field (B $>$ 0.05 T), the electrons motion transverse to the B-field as well as along the B-field is hindered, resulting in a lower electron current to the probe surface. The reduction of the net electron current makes the spherical object less negative. Since ions are assumed to be unmagnetized for the given range of the magnetic field, the role of the magnetic field on the ion current is considered to be negligible. Thus, the electron current determines the surface potential, given by the balance of the electron and ion currents. The value of $V_s$ depends on the magnetic field line density around a spherical object which affects the current to the surface of the object. Therefore, the surface potential is lower (or less negative) for a non-magnetic sphere than for a magnetic sphere in the magnetized plasma.\par
This work highlights the role of the external magnetic field as well as the types of material of the spherical objects (large dust grains) on the surface potential in a low-temperature plasma.
These findings will directly help to estimate the true charges on sub-micron to micron sized dust grains ($r < \lambda_{De}$) in magnetized dusty plasma experiments. It has been confirmed that electron temperature increases as the magnetic field is introduced, which definitely indicates the higher dust charges at lower magnetic field. However, many simulations and experimental works suggest either no changes or less negative charge on the dust grain at low B-field. We expect the reduction in the charge of dust particles ($r < \lambda_{De}$) similar as for spherical probe ($r > \lambda_{De}$) at strong magnetic field. Interestingly, the smaller magnetic or paramagnetic dust grains may acquire a more negative charge on their surface as compared to the non-magnetic dust grains in a magnetized dusty plasma. In the future, our focus will be on the direct or indirect measurement of the charge on dust grains ($r < \lambda_{De}$) in a magnetized dusty plasma to understand the dynamics of the dust grain medium. The reported experimental work may be a motive for researchers to develop an analytical or simulation model to understand the charging mechanism of magnetic and non-magnetic spherical particles in a strongly magnetized plasma. 
 \section{Acknowledgement} 
 This work is supported by the Deutsche Forschungsgemeinschaft (DFG). The authors are grateful to Felix Becker for his assistance in making spherical probes and supporting experiments. The authors are also thankful to M. Kretschmer for the experimental assistance.
\bibliography{biblography}

\begin{thebibliography}{68}%
\makeatletter
\providecommand \@ifxundefined [1]{%
 \@ifx{#1\undefined}
}%
\providecommand \@ifnum [1]{%
 \ifnum #1\expandafter \@firstoftwo
 \else \expandafter \@secondoftwo
 \fi
}%
\providecommand \@ifx [1]{%
 \ifx #1\expandafter \@firstoftwo
 \else \expandafter \@secondoftwo
 \fi
}%
\providecommand \natexlab [1]{#1}%
\providecommand \enquote  [1]{``#1''}%
\providecommand \bibnamefont  [1]{#1}%
\providecommand \bibfnamefont [1]{#1}%
\providecommand \citenamefont [1]{#1}%
\providecommand \href@noop [0]{\@secondoftwo}%
\providecommand \href [0]{\begingroup \@sanitize@url \@href}%
\providecommand \@href[1]{\@@startlink{#1}\@@href}%
\providecommand \@@href[1]{\endgroup#1\@@endlink}%
\providecommand \@sanitize@url [0]{\catcode `\\12\catcode `\$12\catcode
  `\&12\catcode `\#12\catcode `\^12\catcode `\_12\catcode `\%12\relax}%
\providecommand \@@startlink[1]{}%
\providecommand \@@endlink[0]{}%
\providecommand \url  [0]{\begingroup\@sanitize@url \@url }%
\providecommand \@url [1]{\endgroup\@href {#1}{\urlprefix }}%
\providecommand \urlprefix  [0]{URL }%
\providecommand \Eprint [0]{\href }%
\providecommand \doibase [0]{http://dx.doi.org/}%
\providecommand \selectlanguage [0]{\@gobble}%
\providecommand \bibinfo  [0]{\@secondoftwo}%
\providecommand \bibfield  [0]{\@secondoftwo}%
\providecommand \translation [1]{[#1]}%
\providecommand \BibitemOpen [0]{}%
\providecommand \bibitemStop [0]{}%
\providecommand \bibitemNoStop [0]{.\EOS\space}%
\providecommand \EOS [0]{\spacefactor3000\relax}%
\providecommand \BibitemShut  [1]{\csname bibitem#1\endcsname}%
\let\auto@bib@innerbib\@empty
\bibitem [{\citenamefont {Barkan}, \citenamefont {Merlino},\ and\ \citenamefont
  {D'Angelo}(1995)}]{daw2}%
  \BibitemOpen
  \bibfield  {author} {\bibinfo {author} {\bibfnamefont {A.}~\bibnamefont
  {Barkan}}, \bibinfo {author} {\bibfnamefont {R.~L.}\ \bibnamefont {Merlino}},
  \ and\ \bibinfo {author} {\bibfnamefont {N.}~\bibnamefont {D'Angelo}},\
  }\bibfield  {title} {\enquote {\bibinfo {title} {Laboratory observation of
  the dust-acoustic wave mode},}\ }\href@noop {} {\bibfield  {journal}
  {\bibinfo  {journal} {Phys. Plasmas}\ }\textbf {\bibinfo {volume} {2}},\
  \bibinfo {pages} {3563--3565} (\bibinfo {year} {1995})}\BibitemShut {NoStop}%
\bibitem [{\citenamefont {Bandyopadhyay}\ \emph {et~al.}(2008)\citenamefont
  {Bandyopadhyay}, \citenamefont {Prasad}, \citenamefont {Sen},\ and\
  \citenamefont {Kaw}}]{pdasw}%
  \BibitemOpen
  \bibfield  {author} {\bibinfo {author} {\bibfnamefont {P.}~\bibnamefont
  {Bandyopadhyay}}, \bibinfo {author} {\bibfnamefont {G.}~\bibnamefont
  {Prasad}}, \bibinfo {author} {\bibfnamefont {A.}~\bibnamefont {Sen}}, \ and\
  \bibinfo {author} {\bibfnamefont {P.~K.}\ \bibnamefont {Kaw}},\ }\bibfield
  {title} {\enquote {\bibinfo {title} {Experimental study of nonlinear dust
  acoustic solitary waves in a dusty plasma},}\ }\href@noop {} {\bibfield
  {journal} {\bibinfo  {journal} {Phys. Rev. Lett.}\ }\textbf {\bibinfo
  {volume} {101}},\ \bibinfo {pages} {065006} (\bibinfo {year}
  {2008})}\BibitemShut {NoStop}%
\bibitem [{\citenamefont {Choudhary}, \citenamefont {Mukherjee},\ and\
  \citenamefont {Bandyopadhyay}(2016)}]{mangilaldaw}%
  \BibitemOpen
  \bibfield  {author} {\bibinfo {author} {\bibfnamefont {M.}~\bibnamefont
  {Choudhary}}, \bibinfo {author} {\bibfnamefont {S.}~\bibnamefont
  {Mukherjee}}, \ and\ \bibinfo {author} {\bibfnamefont {P.}~\bibnamefont
  {Bandyopadhyay}},\ }\bibfield  {title} {\enquote {\bibinfo {title}
  {Propagation characteristics of dust–acoustic waves in presence of a
  floating cylindrical object in the dc discharge plasma},}\ }\href@noop {}
  {\bibfield  {journal} {\bibinfo  {journal} {Physics of Plasmas}\ }\textbf
  {\bibinfo {volume} {23}},\ \bibinfo {pages} {083705} (\bibinfo {year}
  {2016})}\BibitemShut {NoStop}%
\bibitem [{\citenamefont {Dharodi}, \citenamefont {Tiwari},\ and\ \citenamefont
  {Das}(2014)}]{vikramtsw}%
  \BibitemOpen
  \bibfield  {author} {\bibinfo {author} {\bibfnamefont {V.~S.}\ \bibnamefont
  {Dharodi}}, \bibinfo {author} {\bibfnamefont {S.~K.}\ \bibnamefont {Tiwari}},
  \ and\ \bibinfo {author} {\bibfnamefont {A.}~\bibnamefont {Das}},\ }\bibfield
   {title} {\enquote {\bibinfo {title} {Visco-elastic fluid simulations of
  coherent structures in strongly coupled dusty plasma medium},}\ }\href@noop
  {} {\bibfield  {journal} {\bibinfo  {journal} {Physics of Plasmas}\ }\textbf
  {\bibinfo {volume} {21}},\ \bibinfo {pages} {073705} (\bibinfo {year}
  {2014})}\BibitemShut {NoStop}%
\bibitem [{\citenamefont {Vaulina}\ \emph {et~al.}(2004)\citenamefont
  {Vaulina}, \citenamefont {Samarian}, \citenamefont {Petrov}, \citenamefont
  {James},\ and\ \citenamefont {Melandso}}]{Vaulina2004}%
  \BibitemOpen
  \bibfield  {author} {\bibinfo {author} {\bibfnamefont {O.~S.}\ \bibnamefont
  {Vaulina}}, \bibinfo {author} {\bibfnamefont {A.~A.}\ \bibnamefont
  {Samarian}}, \bibinfo {author} {\bibfnamefont {O.~F.}\ \bibnamefont
  {Petrov}}, \bibinfo {author} {\bibfnamefont {B.}~\bibnamefont {James}}, \
  and\ \bibinfo {author} {\bibfnamefont {F.}~\bibnamefont {Melandso}},\
  }\bibfield  {title} {\enquote {\bibinfo {title} {Formation of vortex dust
  structures in inhomogeneous gas-discharge plasmas},}\ }\href@noop {}
  {\bibfield  {journal} {\bibinfo  {journal} {Plasma Physics Reports}\ }\textbf
  {\bibinfo {volume} {30}},\ \bibinfo {pages} {918--936} (\bibinfo {year}
  {2004})}\BibitemShut {NoStop}%
\bibitem [{\citenamefont {Saitou}\ and\ \citenamefont
  {Ishihara}(2013)}]{satoprl}%
  \BibitemOpen
  \bibfield  {author} {\bibinfo {author} {\bibfnamefont {Y.}~\bibnamefont
  {Saitou}}\ and\ \bibinfo {author} {\bibfnamefont {O.}~\bibnamefont
  {Ishihara}},\ }\bibfield  {title} {\enquote {\bibinfo {title} {Dynamic
  circulation in a complex plasma},}\ }\href@noop {} {\bibfield  {journal}
  {\bibinfo  {journal} {Phys. Rev. Lett.}\ }\textbf {\bibinfo {volume} {111}},\
  \bibinfo {pages} {185003} (\bibinfo {year} {2013})}\BibitemShut {NoStop}%
\bibitem [{\citenamefont {Choudhary}, \citenamefont {Mukherjee},\ and\
  \citenamefont {Bandyopadhyay}(2017)}]{mangilalmultiplevortex}%
  \BibitemOpen
  \bibfield  {author} {\bibinfo {author} {\bibfnamefont {M.}~\bibnamefont
  {Choudhary}}, \bibinfo {author} {\bibfnamefont {S.}~\bibnamefont
  {Mukherjee}}, \ and\ \bibinfo {author} {\bibfnamefont {P.}~\bibnamefont
  {Bandyopadhyay}},\ }\bibfield  {title} {\enquote {\bibinfo {title}
  {Experimental observation of self excited co-rotating multiple vortices in a
  dusty plasma with inhomogeneous plasma background},}\ }\href@noop {}
  {\bibfield  {journal} {\bibinfo  {journal} {Physics of Plasmas}\ }\textbf
  {\bibinfo {volume} {24}},\ \bibinfo {pages} {033703} (\bibinfo {year}
  {2017})}\BibitemShut {NoStop}%
\bibitem [{\citenamefont {Law}\ \emph {et~al.}(1998)\citenamefont {Law},
  \citenamefont {Steel}, \citenamefont {Annaratone},\ and\ \citenamefont
  {Allen}}]{probeinducedcirculation}%
  \BibitemOpen
  \bibfield  {author} {\bibinfo {author} {\bibfnamefont {D.~A.}\ \bibnamefont
  {Law}}, \bibinfo {author} {\bibfnamefont {W.~H.}\ \bibnamefont {Steel}},
  \bibinfo {author} {\bibfnamefont {B.~M.}\ \bibnamefont {Annaratone}}, \ and\
  \bibinfo {author} {\bibfnamefont {J.~E.}\ \bibnamefont {Allen}},\ }\bibfield
  {title} {\enquote {\bibinfo {title} {Probe-induced particle circulation in a
  plasma crystal},}\ }\href@noop {} {\bibfield  {journal} {\bibinfo  {journal}
  {Phys. Rev. Lett.}\ }\textbf {\bibinfo {volume} {80}},\ \bibinfo {pages}
  {4189--4192} (\bibinfo {year} {1998})}\BibitemShut {NoStop}%
\bibitem [{\citenamefont {Choudhary}, \citenamefont {Mukherjee},\ and\
  \citenamefont {Bandyopadhyay}(2018)}]{mangilallargeaspect}%
  \BibitemOpen
  \bibfield  {author} {\bibinfo {author} {\bibfnamefont {M.}~\bibnamefont
  {Choudhary}}, \bibinfo {author} {\bibfnamefont {S.}~\bibnamefont
  {Mukherjee}}, \ and\ \bibinfo {author} {\bibfnamefont {P.}~\bibnamefont
  {Bandyopadhyay}},\ }\bibfield  {title} {\enquote {\bibinfo {title}
  {Collective dynamics of large aspect ratio dusty plasma in an inhomogeneous
  plasma background: Formation of the co-rotating vortex series},}\ }\href@noop
  {} {\bibfield  {journal} {\bibinfo  {journal} {Physics of Plasmas}\ }\textbf
  {\bibinfo {volume} {25}},\ \bibinfo {pages} {023704} (\bibinfo {year}
  {2018})}\BibitemShut {NoStop}%
\bibitem [{\citenamefont {Choudhary}\ \emph {et~al.}(2019)\citenamefont
  {Choudhary}, \citenamefont {Bergert}, \citenamefont {Mitic},\ and\
  \citenamefont {Thoma}}]{mangimagneticrotation}%
  \BibitemOpen
  \bibfield  {author} {\bibinfo {author} {\bibfnamefont {M.}~\bibnamefont
  {Choudhary}}, \bibinfo {author} {\bibfnamefont {R.}~\bibnamefont {Bergert}},
  \bibinfo {author} {\bibfnamefont {S.}~\bibnamefont {Mitic}}, \ and\ \bibinfo
  {author} {\bibfnamefont {M.~H.}\ \bibnamefont {Thoma}},\ }\bibfield  {title}
  {\enquote {\bibinfo {title} {3-dimensional dusty plasma in a strong magnetic
  field: Observation of rotating dust tori},}\ }\href@noop {} {\bibfield
  {journal} {\bibinfo  {journal} {https://arxiv.org/abs/1910.07846}\ }\textbf
  {\bibinfo {volume} {0}},\ \bibinfo {pages} {0} (\bibinfo {year}
  {2019})}\BibitemShut {NoStop}%
\bibitem [{\citenamefont {Goertz}(1989)}]{goertzdustysolarsystem}%
  \BibitemOpen
  \bibfield  {author} {\bibinfo {author} {\bibfnamefont {C.~K.}\ \bibnamefont
  {Goertz}},\ }\bibfield  {title} {\enquote {\bibinfo {title} {Dusty plasmas in
  the solar system},}\ }\href@noop {} {\bibfield  {journal} {\bibinfo
  {journal} {Reviews of Geophysics}\ }\textbf {\bibinfo {volume} {27}},\
  \bibinfo {pages} {271--292} (\bibinfo {year} {1989})}\BibitemShut {NoStop}%
\bibitem [{\citenamefont {Goertz}(1984)}]{geortzspokes2}%
  \BibitemOpen
  \bibfield  {author} {\bibinfo {author} {\bibfnamefont {C.}~\bibnamefont
  {Goertz}},\ }\bibfield  {title} {\enquote {\bibinfo {title} {Formation of
  saturn's spokes},}\ }\href@noop {} {\bibfield  {journal} {\bibinfo  {journal}
  {Advances in Space Research}\ }\textbf {\bibinfo {volume} {4}},\ \bibinfo
  {pages} {137--141} (\bibinfo {year} {1984})}\BibitemShut {NoStop}%
\bibitem [{\citenamefont {Mendis}\ and\ \citenamefont
  {Rosenberg}(1994)}]{cosmicdustymendis}%
  \BibitemOpen
  \bibfield  {author} {\bibinfo {author} {\bibfnamefont {D.~A.}\ \bibnamefont
  {Mendis}}\ and\ \bibinfo {author} {\bibfnamefont {M.}~\bibnamefont
  {Rosenberg}},\ }\bibfield  {title} {\enquote {\bibinfo {title} {Cosmic dusty
  plasma},}\ }\href@noop {} {\bibfield  {journal} {\bibinfo  {journal} {Annu.
  Rev. Astron. Astrophys.}\ }\textbf {\bibinfo {volume} {32}},\ \bibinfo
  {pages} {419--63} (\bibinfo {year} {1994})}\BibitemShut {NoStop}%
\bibitem [{\citenamefont {Selwyn}, \citenamefont {Heidenreich},\ and\
  \citenamefont {Haller}(1991)}]{selwynprocessing1}%
  \BibitemOpen
  \bibfield  {author} {\bibinfo {author} {\bibfnamefont {G.~S.}\ \bibnamefont
  {Selwyn}}, \bibinfo {author} {\bibfnamefont {J.~E.}\ \bibnamefont
  {Heidenreich}}, \ and\ \bibinfo {author} {\bibfnamefont {K.~L.}\ \bibnamefont
  {Haller}},\ }\bibfield  {title} {\enquote {\bibinfo {title} {Rastered laser
  light scattering studies during plasma processing: Particle contamination
  trapping phenomena},}\ }\href@noop {} {\bibfield  {journal} {\bibinfo
  {journal} {J. Vac. Sci. Technol. A}\ }\textbf {\bibinfo {volume} {9}},\
  \bibinfo {pages} {2817--2824} (\bibinfo {year} {1991})}\BibitemShut {NoStop}%
\bibitem [{\citenamefont {Watanabe}(1997)}]{watanableprocessingplasma2}%
  \BibitemOpen
  \bibfield  {author} {\bibinfo {author} {\bibfnamefont {Y.}~\bibnamefont
  {Watanabe}},\ }\bibfield  {title} {\enquote {\bibinfo {title} {Dust phenomena
  in processing plasmas},}\ }\href@noop {} {\bibfield  {journal} {\bibinfo
  {journal} {Plasma Phys. Control. Fusion}\ }\textbf {\bibinfo {volume} {39}},\
  \bibinfo {pages} {A59--A72} (\bibinfo {year} {1997})}\BibitemShut {NoStop}%
\bibitem [{\citenamefont {Winter}(2000)}]{winterfusiondust}%
  \BibitemOpen
  \bibfield  {author} {\bibinfo {author} {\bibfnamefont {J.}~\bibnamefont
  {Winter}},\ }\bibfield  {title} {\enquote {\bibinfo {title} {Dust: A new
  challenge in nuclear fusion research?}}\ }\href@noop {} {\bibfield  {journal}
  {\bibinfo  {journal} {Phys. Plasmas}\ }\textbf {\bibinfo {volume} {7}},\
  \bibinfo {pages} {3862--3866} (\bibinfo {year} {2000})}\BibitemShut {NoStop}%
\bibitem [{\citenamefont {Löwen}\ \emph {et~al.}(2011)\citenamefont {Löwen},
  \citenamefont {Royall}, \citenamefont {Ivlev},\ and\ \citenamefont
  {Morfill}}]{colloidaldustyplasma}%
  \BibitemOpen
  \bibfield  {author} {\bibinfo {author} {\bibfnamefont {H.}~\bibnamefont
  {Löwen}}, \bibinfo {author} {\bibfnamefont {C.~P.}\ \bibnamefont {Royall}},
  \bibinfo {author} {\bibfnamefont {A.}~\bibnamefont {Ivlev}}, \ and\ \bibinfo
  {author} {\bibfnamefont {G.~E.}\ \bibnamefont {Morfill}},\ }\bibfield
  {title} {\enquote {\bibinfo {title} {Charged colloidal suspensions and their
  link to complex plasmas},}\ }\href@noop {} {\bibfield  {journal} {\bibinfo
  {journal} {AIP Conference Proceedings}\ }\textbf {\bibinfo {volume} {1397}},\
  \bibinfo {pages} {201--210} (\bibinfo {year} {2011})}\BibitemShut {NoStop}%
\bibitem [{\citenamefont {Barkan}, \citenamefont {D'Angelo},\ and\
  \citenamefont {Merlino}(1994)}]{Charging}%
  \BibitemOpen
  \bibfield  {author} {\bibinfo {author} {\bibfnamefont {A.}~\bibnamefont
  {Barkan}}, \bibinfo {author} {\bibfnamefont {N.}~\bibnamefont {D'Angelo}}, \
  and\ \bibinfo {author} {\bibfnamefont {R.~L.}\ \bibnamefont {Merlino}},\
  }\bibfield  {title} {\enquote {\bibinfo {title} {Charging of dust grains in a
  plasma},}\ }\href@noop {} {\bibfield  {journal} {\bibinfo  {journal} {Phys.
  Rev. Lett.}\ }\textbf {\bibinfo {volume} {73}},\ \bibinfo {pages}
  {3093--3096} (\bibinfo {year} {1994})}\BibitemShut {NoStop}%
\bibitem [{\citenamefont {Goree}(1994)}]{goreedustcharging1}%
  \BibitemOpen
  \bibfield  {author} {\bibinfo {author} {\bibfnamefont {J.}~\bibnamefont
  {Goree}},\ }\bibfield  {title} {\enquote {\bibinfo {title} {Charging of
  particles in a plasma},}\ }\href@noop {} {\bibfield  {journal} {\bibinfo
  {journal} {Plasma Sources Sci. Technol.}\ }\textbf {\bibinfo {volume} {3}},\
  \bibinfo {pages} {400--406} (\bibinfo {year} {1994})}\BibitemShut {NoStop}%
\bibitem [{\citenamefont {Wu}\ and\ \citenamefont
  {Miller}(1990)}]{jwangdustcharge1}%
  \BibitemOpen
  \bibfield  {author} {\bibinfo {author} {\bibfnamefont {J.~J.}\ \bibnamefont
  {Wu}}\ and\ \bibinfo {author} {\bibfnamefont {R.~J.}\ \bibnamefont
  {Miller}},\ }\bibfield  {title} {\enquote {\bibinfo {title} {Measurements of
  charge on submicron particles generated in a sputtering process},}\
  }\href@noop {} {\bibfield  {journal} {\bibinfo  {journal} {J. Appl. Physics}\
  }\textbf {\bibinfo {volume} {67}},\ \bibinfo {pages} {1051--1054} (\bibinfo
  {year} {1990})}\BibitemShut {NoStop}%
\bibitem [{\citenamefont {Walch}, \citenamefont {Horanyi},\ and\ \citenamefont
  {Robertson}(1994)}]{walchdustcharge2}%
  \BibitemOpen
  \bibfield  {author} {\bibinfo {author} {\bibfnamefont {B.}~\bibnamefont
  {Walch}}, \bibinfo {author} {\bibfnamefont {M.}~\bibnamefont {Horanyi}}, \
  and\ \bibinfo {author} {\bibfnamefont {S.}~\bibnamefont {Robertson}},\
  }\bibfield  {title} {\enquote {\bibinfo {title} {Measurement of the charging
  of individual dust grains in a plasma},}\ }\href@noop {} {\bibfield
  {journal} {\bibinfo  {journal} {IEEE Transactions on Plasma Science}\
  }\textbf {\bibinfo {volume} {22}},\ \bibinfo {pages} {97--102} (\bibinfo
  {year} {1994})}\BibitemShut {NoStop}%
\bibitem [{\citenamefont {Khrapak}\ \emph
  {et~al.}(2005{\natexlab{a}})\citenamefont {Khrapak}, \citenamefont
  {Ratynskaia}, \citenamefont {Zobnin}, \citenamefont {Usachev}, \citenamefont
  {Yaroshenko}, \citenamefont {Thoma}, \citenamefont {Kretschmer},
  \citenamefont {H\"ofner}, \citenamefont {Morfill}, \citenamefont {Petrov},\
  and\ \citenamefont {Fortov}}]{particlechargemarkus1}%
  \BibitemOpen
  \bibfield  {author} {\bibinfo {author} {\bibfnamefont {S.~A.}\ \bibnamefont
  {Khrapak}}, \bibinfo {author} {\bibfnamefont {S.~V.}\ \bibnamefont
  {Ratynskaia}}, \bibinfo {author} {\bibfnamefont {A.~V.}\ \bibnamefont
  {Zobnin}}, \bibinfo {author} {\bibfnamefont {A.~D.}\ \bibnamefont {Usachev}},
  \bibinfo {author} {\bibfnamefont {V.~V.}\ \bibnamefont {Yaroshenko}},
  \bibinfo {author} {\bibfnamefont {M.~H.}\ \bibnamefont {Thoma}}, \bibinfo
  {author} {\bibfnamefont {M.}~\bibnamefont {Kretschmer}}, \bibinfo {author}
  {\bibfnamefont {H.}~\bibnamefont {H\"ofner}}, \bibinfo {author}
  {\bibfnamefont {G.~E.}\ \bibnamefont {Morfill}}, \bibinfo {author}
  {\bibfnamefont {O.~F.}\ \bibnamefont {Petrov}}, \ and\ \bibinfo {author}
  {\bibfnamefont {V.~E.}\ \bibnamefont {Fortov}},\ }\bibfield  {title}
  {\enquote {\bibinfo {title} {Particle charge in the bulk of gas
  discharges},}\ }\href@noop {} {\bibfield  {journal} {\bibinfo  {journal}
  {Phys. Rev. E}\ }\textbf {\bibinfo {volume} {72}},\ \bibinfo {pages} {016406}
  (\bibinfo {year} {2005}{\natexlab{a}})}\BibitemShut {NoStop}%
\bibitem [{\citenamefont {Mott-Smith}\ and\ \citenamefont
  {Langmuir}(1926)}]{mottsmitomltheory1}%
  \BibitemOpen
  \bibfield  {author} {\bibinfo {author} {\bibfnamefont {H.~M.}\ \bibnamefont
  {Mott-Smith}}\ and\ \bibinfo {author} {\bibfnamefont {I.}~\bibnamefont
  {Langmuir}},\ }\bibfield  {title} {\enquote {\bibinfo {title} {The theory of
  collectors in gaseous discharges},}\ }\href@noop {} {\bibfield  {journal}
  {\bibinfo  {journal} {Phys. Rev.}\ }\textbf {\bibinfo {volume} {28}},\
  \bibinfo {pages} {727--763} (\bibinfo {year} {1926})}\BibitemShut {NoStop}%
\bibitem [{\citenamefont {Allen}(1992)}]{allenomltheory2}%
  \BibitemOpen
  \bibfield  {author} {\bibinfo {author} {\bibfnamefont {J.~E.}\ \bibnamefont
  {Allen}},\ }\bibfield  {title} {\enquote {\bibinfo {title} {Probe theory -
  the orbital motion approach},}\ }\href@noop {} {\bibfield  {journal}
  {\bibinfo  {journal} {Physica Scripta}\ }\textbf {\bibinfo {volume} {45}},\
  \bibinfo {pages} {497} (\bibinfo {year} {1992})}\BibitemShut {NoStop}%
\bibitem [{\citenamefont {Willis}\ \emph {et~al.}(2010)\citenamefont {Willis},
  \citenamefont {Coppins}, \citenamefont {Bacharis},\ and\ \citenamefont
  {Allen}}]{willisfloatingpotential2}%
  \BibitemOpen
  \bibfield  {author} {\bibinfo {author} {\bibfnamefont {C.~T.~N.}\
  \bibnamefont {Willis}}, \bibinfo {author} {\bibfnamefont {M.}~\bibnamefont
  {Coppins}}, \bibinfo {author} {\bibfnamefont {M.}~\bibnamefont {Bacharis}}, \
  and\ \bibinfo {author} {\bibfnamefont {J.~E.}\ \bibnamefont {Allen}},\
  }\bibfield  {title} {\enquote {\bibinfo {title} {The effect of dust grain
  size on the floating potential of dust in a collisionless plasma},}\
  }\href@noop {} {\bibfield  {journal} {\bibinfo  {journal} {Plasma Sources
  Sci. Technol.}\ }\textbf {\bibinfo {volume} {19}},\ \bibinfo {pages} {065022}
  (\bibinfo {year} {2010})}\BibitemShut {NoStop}%
\bibitem [{\citenamefont {Morfill}\ and\ \citenamefont
  {Ivlev}(2009)}]{morfilldusty1}%
  \BibitemOpen
  \bibfield  {author} {\bibinfo {author} {\bibfnamefont {G.~E.}\ \bibnamefont
  {Morfill}}\ and\ \bibinfo {author} {\bibfnamefont {A.~V.}\ \bibnamefont
  {Ivlev}},\ }\bibfield  {title} {\enquote {\bibinfo {title} {Complex plasmas:
  An interdisciplinary research field},}\ }\href@noop {} {\bibfield  {journal}
  {\bibinfo  {journal} {Rev. Mod. Phys.}\ }\textbf {\bibinfo {volume} {81}},\
  \bibinfo {pages} {1353--1404} (\bibinfo {year} {2009})}\BibitemShut {NoStop}%
\bibitem [{\citenamefont {Bonitz}, \citenamefont {Henning},\ and\ \citenamefont
  {Block}(2010)}]{bonitzdusty2}%
  \BibitemOpen
  \bibfield  {author} {\bibinfo {author} {\bibfnamefont {M.}~\bibnamefont
  {Bonitz}}, \bibinfo {author} {\bibfnamefont {C.}~\bibnamefont {Henning}}, \
  and\ \bibinfo {author} {\bibfnamefont {D.}~\bibnamefont {Block}},\ }\bibfield
   {title} {\enquote {\bibinfo {title} {Complex plasmas: a laboratory for
  strong correlations},}\ }\href@noop {} {\bibfield  {journal} {\bibinfo
  {journal} {Reports on Progress in Physics}\ }\textbf {\bibinfo {volume}
  {73}},\ \bibinfo {pages} {066501} (\bibinfo {year} {2010})}\BibitemShut
  {NoStop}%
\bibitem [{\citenamefont {Tsytovich}, \citenamefont {Sato},\ and\ \citenamefont
  {Morfill}(2003)}]{tsytovichdustcharge1}%
  \BibitemOpen
  \bibfield  {author} {\bibinfo {author} {\bibfnamefont {V.~N.}\ \bibnamefont
  {Tsytovich}}, \bibinfo {author} {\bibfnamefont {N.}~\bibnamefont {Sato}}, \
  and\ \bibinfo {author} {\bibfnamefont {G.~E.}\ \bibnamefont {Morfill}},\
  }\bibfield  {title} {\enquote {\bibinfo {title} {Note on the charging and
  spinning of dust particles in complex plasmas in a strong magnetic field},}\
  }\href@noop {} {\bibfield  {journal} {\bibinfo  {journal} {New Journal of
  Physics}\ }\textbf {\bibinfo {volume} {5}},\ \bibinfo {pages} {43--43}
  (\bibinfo {year} {2003})}\BibitemShut {NoStop}%
\bibitem [{\citenamefont {Lange}(2016)}]{langefloatinginmagnetized}%
  \BibitemOpen
  \bibfield  {author} {\bibinfo {author} {\bibfnamefont {D.}~\bibnamefont
  {Lange}},\ }\bibfield  {title} {\enquote {\bibinfo {title} {Floating surface
  potential of spherical dust grains in magnetized plasmas},}\ }\href@noop {}
  {\bibfield  {journal} {\bibinfo  {journal} {Journal of Plasma Physics}\
  }\textbf {\bibinfo {volume} {82}},\ \bibinfo {pages} {905820101} (\bibinfo
  {year} {2016})}\BibitemShut {NoStop}%
\bibitem [{\citenamefont {Patacchini}, \citenamefont {Hutchinson},\ and\
  \citenamefont {Lapenta}(2007)}]{dustcurrent}%
  \BibitemOpen
  \bibfield  {author} {\bibinfo {author} {\bibfnamefont {L.}~\bibnamefont
  {Patacchini}}, \bibinfo {author} {\bibfnamefont {I.~H.}\ \bibnamefont
  {Hutchinson}}, \ and\ \bibinfo {author} {\bibfnamefont {G.}~\bibnamefont
  {Lapenta}},\ }\bibfield  {title} {\enquote {\bibinfo {title} {Electron
  collection by a negatively charged sphere in a collisionless
  magnetoplasma},}\ }\href@noop {} {\bibfield  {journal} {\bibinfo  {journal}
  {Physics of Plasmas}\ }\textbf {\bibinfo {volume} {14}},\ \bibinfo {pages}
  {062111} (\bibinfo {year} {2007})}\BibitemShut {NoStop}%
\bibitem [{\citenamefont {{Kodanova}}\ \emph {et~al.}(2019)\citenamefont
  {{Kodanova}}, \citenamefont {{Bastykova}}, \citenamefont {{Ramazanov}},
  \citenamefont {{Nigmetova}}, \citenamefont {{Maiorov}},\ and\ \citenamefont
  {{Moldabekov}}}]{kodanovadustchargeb}%
  \BibitemOpen
  \bibfield  {author} {\bibinfo {author} {\bibfnamefont {S.~K.}\ \bibnamefont
  {{Kodanova}}}, \bibinfo {author} {\bibfnamefont {N.~K.}\ \bibnamefont
  {{Bastykova}}}, \bibinfo {author} {\bibfnamefont {T.~S.}\ \bibnamefont
  {{Ramazanov}}}, \bibinfo {author} {\bibfnamefont {G.~N.}\ \bibnamefont
  {{Nigmetova}}}, \bibinfo {author} {\bibfnamefont {S.~A.}\ \bibnamefont
  {{Maiorov}}}, \ and\ \bibinfo {author} {\bibfnamefont {Z.~A.}\ \bibnamefont
  {{Moldabekov}}},\ }\bibfield  {title} {\enquote {\bibinfo {title} {Charging
  of a dust particle in a magnetized gas discharge plasma},}\ }\href@noop {}
  {\bibfield  {journal} {\bibinfo  {journal} {IEEE Transactions on Plasma
  Science}\ }\textbf {\bibinfo {volume} {47}},\ \bibinfo {pages} {3052--3056}
  (\bibinfo {year} {2019})}\BibitemShut {NoStop}%
\bibitem [{\citenamefont {Yukihiro}\ \emph {et~al.}(2009)\citenamefont
  {Yukihiro}, \citenamefont {Gakushi}, \citenamefont {Takatoshi},\ and\
  \citenamefont {Osamu}}]{tomitadustchargingwithmagneticfield}%
  \BibitemOpen
  \bibfield  {author} {\bibinfo {author} {\bibfnamefont {T.}~\bibnamefont
  {Yukihiro}}, \bibinfo {author} {\bibfnamefont {K.}~\bibnamefont {Gakushi}},
  \bibinfo {author} {\bibfnamefont {Y.}~\bibnamefont {Takatoshi}}, \ and\
  \bibinfo {author} {\bibfnamefont {I.}~\bibnamefont {Osamu}},\ }\bibfield
  {title} {\enquote {\bibinfo {title} {Charging of dust particles in magnetic
  field},}\ }\href@noop {} {\bibfield  {journal} {\bibinfo  {journal} {J.
  Plasma Fusion Res. SERIES}\ }\textbf {\bibinfo {volume} {8}},\ \bibinfo
  {pages} {273--276} (\bibinfo {year} {2009})}\BibitemShut {NoStop}%
\bibitem [{\citenamefont {Kalita}\ \emph {et~al.}(2015)\citenamefont {Kalita},
  \citenamefont {Kakati}, \citenamefont {Saikia}, \citenamefont
  {Bandyopadhyay},\ and\ \citenamefont {Kausik}}]{kalita}%
  \BibitemOpen
  \bibfield  {author} {\bibinfo {author} {\bibfnamefont {D.}~\bibnamefont
  {Kalita}}, \bibinfo {author} {\bibfnamefont {B.}~\bibnamefont {Kakati}},
  \bibinfo {author} {\bibfnamefont {B.~K.}\ \bibnamefont {Saikia}}, \bibinfo
  {author} {\bibfnamefont {M.}~\bibnamefont {Bandyopadhyay}}, \ and\ \bibinfo
  {author} {\bibfnamefont {S.~S.}\ \bibnamefont {Kausik}},\ }\bibfield  {title}
  {\enquote {\bibinfo {title} {Effect of magnetic field on dust charging and
  corresponding probe measurement},}\ }\href@noop {} {\bibfield  {journal}
  {\bibinfo  {journal} {Physics of Plasmas}\ }\textbf {\bibinfo {volume}
  {22}},\ \bibinfo {pages} {113704} (\bibinfo {year} {2015})}\BibitemShut
  {NoStop}%
\bibitem [{\citenamefont {Tadsen}, \citenamefont {Greiner},\ and\ \citenamefont
  {Piel}(2018)}]{melzermagnetizeddusty}%
  \BibitemOpen
  \bibfield  {author} {\bibinfo {author} {\bibfnamefont {B.}~\bibnamefont
  {Tadsen}}, \bibinfo {author} {\bibfnamefont {F.}~\bibnamefont {Greiner}}, \
  and\ \bibinfo {author} {\bibfnamefont {A.}~\bibnamefont {Piel}},\ }\bibfield
  {title} {\enquote {\bibinfo {title} {Probing a dusty magnetized plasma with
  self-excited dust-density waves},}\ }\href@noop {} {\bibfield  {journal}
  {\bibinfo  {journal} {Phys. Rev. E}\ }\textbf {\bibinfo {volume} {97}},\
  \bibinfo {pages} {033203} (\bibinfo {year} {2018})}\BibitemShut {NoStop}%
\bibitem [{\citenamefont {Melzer}\ \emph {et~al.}(2019)\citenamefont {Melzer},
  \citenamefont {Krüger}, \citenamefont {Schütt},\ and\ \citenamefont
  {Mulsow}}]{melzerdustchargeb}%
  \BibitemOpen
  \bibfield  {author} {\bibinfo {author} {\bibfnamefont {A.}~\bibnamefont
  {Melzer}}, \bibinfo {author} {\bibfnamefont {H.}~\bibnamefont {Krüger}},
  \bibinfo {author} {\bibfnamefont {S.}~\bibnamefont {Schütt}}, \ and\
  \bibinfo {author} {\bibfnamefont {M.}~\bibnamefont {Mulsow}},\ }\bibfield
  {title} {\enquote {\bibinfo {title} {Finite dust clusters under strong
  magnetic fields},}\ }\href@noop {} {\bibfield  {journal} {\bibinfo  {journal}
  {Physics of Plasmas}\ }\textbf {\bibinfo {volume} {26}},\ \bibinfo {pages}
  {093702} (\bibinfo {year} {2019})}\BibitemShut {NoStop}%
\bibitem [{\citenamefont {Puttscher}\ and\ \citenamefont
  {Melzer}(2014)}]{paramagneticdust}%
  \BibitemOpen
  \bibfield  {author} {\bibinfo {author} {\bibfnamefont {M.}~\bibnamefont
  {Puttscher}}\ and\ \bibinfo {author} {\bibfnamefont {A.}~\bibnamefont
  {Melzer}},\ }\bibfield  {title} {\enquote {\bibinfo {title} {Paramagnetic
  dust particles in rf-plasmas with weak external magnetic fields},}\
  }\href@noop {} {\bibfield  {journal} {\bibinfo  {journal} {New Journal of
  Physics}\ }\textbf {\bibinfo {volume} {16}},\ \bibinfo {pages} {043026}
  (\bibinfo {year} {2014})}\BibitemShut {NoStop}%
\bibitem [{\citenamefont {Schwabe}\ \emph {et~al.}(2011)\citenamefont
  {Schwabe}, \citenamefont {Konopka}, \citenamefont {Bandyopadhyay},\ and\
  \citenamefont {Morfill}}]{kanopkamagnetized}%
  \BibitemOpen
  \bibfield  {author} {\bibinfo {author} {\bibfnamefont {M.}~\bibnamefont
  {Schwabe}}, \bibinfo {author} {\bibfnamefont {U.}~\bibnamefont {Konopka}},
  \bibinfo {author} {\bibfnamefont {P.}~\bibnamefont {Bandyopadhyay}}, \ and\
  \bibinfo {author} {\bibfnamefont {G.~E.}\ \bibnamefont {Morfill}},\
  }\bibfield  {title} {\enquote {\bibinfo {title} {Pattern formation in a
  complex plasma in high magnetic fields},}\ }\href@noop {} {\bibfield
  {journal} {\bibinfo  {journal} {Phys. Rev. Lett.}\ }\textbf {\bibinfo
  {volume} {106}},\ \bibinfo {pages} {215004} (\bibinfo {year}
  {2011})}\BibitemShut {NoStop}%
\bibitem [{\citenamefont {Sheehan}\ and\ \citenamefont
  {Hershkowitz}(2011)}]{emissivesheehan}%
  \BibitemOpen
  \bibfield  {author} {\bibinfo {author} {\bibfnamefont {J.~P.}\ \bibnamefont
  {Sheehan}}\ and\ \bibinfo {author} {\bibfnamefont {N.}~\bibnamefont
  {Hershkowitz}},\ }\bibfield  {title} {\enquote {\bibinfo {title} {Emissive
  probes},}\ }\href@noop {} {\bibfield  {journal} {\bibinfo  {journal} {Plasma
  Sources Sci. Technol.}\ }\textbf {\bibinfo {volume} {20}},\ \bibinfo {pages}
  {063001} (\bibinfo {year} {2011})}\BibitemShut {NoStop}%
\bibitem [{\citenamefont {Balan}\ \emph {et~al.}(2003)\citenamefont {Balan},
  \citenamefont {Schrittwieser}, \citenamefont {Ioniţă}, \citenamefont
  {Cabral}, \citenamefont {Figueiredo}, \citenamefont {Fernandes},
  \citenamefont {Varandas}, \citenamefont {Adámek}, \citenamefont {Hron},
  \citenamefont {Stöckel}, \citenamefont {Martines}, \citenamefont {Tichý},\
  and\ \citenamefont {Van~Oost}}]{balanemissiveprobetokamak}%
  \BibitemOpen
  \bibfield  {author} {\bibinfo {author} {\bibfnamefont {P.}~\bibnamefont
  {Balan}}, \bibinfo {author} {\bibfnamefont {R.}~\bibnamefont
  {Schrittwieser}}, \bibinfo {author} {\bibfnamefont {C.}~\bibnamefont
  {Ioniţă}}, \bibinfo {author} {\bibfnamefont {J.~A.}\ \bibnamefont
  {Cabral}}, \bibinfo {author} {\bibfnamefont {H.~F.~C.}\ \bibnamefont
  {Figueiredo}}, \bibinfo {author} {\bibfnamefont {H.}~\bibnamefont
  {Fernandes}}, \bibinfo {author} {\bibfnamefont {C.}~\bibnamefont {Varandas}},
  \bibinfo {author} {\bibfnamefont {J.}~\bibnamefont {Adámek}}, \bibinfo
  {author} {\bibfnamefont {M.}~\bibnamefont {Hron}}, \bibinfo {author}
  {\bibfnamefont {J.}~\bibnamefont {Stöckel}}, \bibinfo {author}
  {\bibfnamefont {E.}~\bibnamefont {Martines}}, \bibinfo {author}
  {\bibfnamefont {M.}~\bibnamefont {Tichý}}, \ and\ \bibinfo {author}
  {\bibfnamefont {G.}~\bibnamefont {Van~Oost}},\ }\bibfield  {title} {\enquote
  {\bibinfo {title} {Emissive probe measurements of plasma potential
  fluctuations in the edge plasma regions of tokamaks},}\ }\href@noop {}
  {\bibfield  {journal} {\bibinfo  {journal} {Rev. Sci. Instrum.}\ }\textbf
  {\bibinfo {volume} {74}},\ \bibinfo {pages} {1583--1587} (\bibinfo {year}
  {2003})}\BibitemShut {NoStop}%
\bibitem [{\citenamefont {Fujita}\ \emph {et~al.}(1980)\citenamefont {Fujita},
  \citenamefont {Nowak}, \citenamefont {Hoegger},\ and\ \citenamefont
  {Schneider}}]{hitujaemmisiveprobe}%
  \BibitemOpen
  \bibfield  {author} {\bibinfo {author} {\bibfnamefont {H.}~\bibnamefont
  {Fujita}}, \bibinfo {author} {\bibfnamefont {S.}~\bibnamefont {Nowak}},
  \bibinfo {author} {\bibfnamefont {B.}~\bibnamefont {Hoegger}}, \ and\
  \bibinfo {author} {\bibfnamefont {H.}~\bibnamefont {Schneider}},\ }\bibfield
  {title} {\enquote {\bibinfo {title} {Potential measurements by an emissive
  probe in a magnetized plasma},}\ }\href@noop {} {\bibfield  {journal}
  {\bibinfo  {journal} {Physics Letters A}\ }\textbf {\bibinfo {volume} {78}},\
  \bibinfo {pages} {263--265} (\bibinfo {year} {1980})}\BibitemShut {NoStop}%
\bibitem [{\citenamefont {Bradley}, \citenamefont {Thompson},\ and\
  \citenamefont {Gonzalvo}(2001)}]{bradelyemmisiveprobeinB}%
  \BibitemOpen
  \bibfield  {author} {\bibinfo {author} {\bibfnamefont {J.~W.}\ \bibnamefont
  {Bradley}}, \bibinfo {author} {\bibfnamefont {S.}~\bibnamefont {Thompson}}, \
  and\ \bibinfo {author} {\bibfnamefont {Y.~A.}\ \bibnamefont {Gonzalvo}},\
  }\bibfield  {title} {\enquote {\bibinfo {title} {Measurement of the plasma
  potential in a magnetron discharge and the prediction of the electron drift
  speeds},}\ }\href@noop {} {\bibfield  {journal} {\bibinfo  {journal} {Plasma
  Sources Sci. Technol.}\ }\textbf {\bibinfo {volume} {10}},\ \bibinfo {pages}
  {490} (\bibinfo {year} {2001})}\BibitemShut {NoStop}%
\bibitem [{\citenamefont {Choudhary}(2017)}]{mangilalthesis}%
  \BibitemOpen
  \bibfield  {author} {\bibinfo {author} {\bibfnamefont {M.}~\bibnamefont
  {Choudhary}},\ }\emph {\bibinfo {title} {Experimental Studies on Collective
  Phenomena in Dusty Plasmas}},\ \href@noop {} {Ph.D. thesis},\ \bibinfo
  {school} {Homi Bhabha National Institute} (\bibinfo {year}
  {2017})\BibitemShut {NoStop}%
\bibitem [{\citenamefont {Schrittwieser}\ \emph {et~al.}(2005)\citenamefont
  {Schrittwieser}, \citenamefont {Ionita}, \citenamefont {Balan}, \citenamefont
  {Varandas}, \citenamefont {Figueiredo}, \citenamefont {Stoeckel},
  \citenamefont {Adamek}, \citenamefont {Hron}, \citenamefont {Ryszawy},
  \citenamefont {M}, \citenamefont {Martines}, \citenamefont {Oost},
  \citenamefont {Klinger}, ,\ and\ \citenamefont
  {Madani}}]{emissiveprobescrutwiser}%
  \BibitemOpen
  \bibfield  {author} {\bibinfo {author} {\bibfnamefont {R.}~\bibnamefont
  {Schrittwieser}}, \bibinfo {author} {\bibfnamefont {C.}~\bibnamefont
  {Ionita}}, \bibinfo {author} {\bibfnamefont {P.~C.}\ \bibnamefont {Balan}},
  \bibinfo {author} {\bibfnamefont {C.~A.}\ \bibnamefont {Varandas}}, \bibinfo
  {author} {\bibfnamefont {H.~F.}\ \bibnamefont {Figueiredo}}, \bibinfo
  {author} {\bibfnamefont {J.}~\bibnamefont {Stoeckel}}, \bibinfo {author}
  {\bibfnamefont {J.}~\bibnamefont {Adamek}}, \bibinfo {author} {\bibfnamefont
  {M.}~\bibnamefont {Hron}}, \bibinfo {author} {\bibfnamefont {J.}~\bibnamefont
  {Ryszawy}}, \bibinfo {author} {\bibfnamefont {T.}~\bibnamefont {M}}, \bibinfo
  {author} {\bibfnamefont {E.}~\bibnamefont {Martines}}, \bibinfo {author}
  {\bibfnamefont {G.~V.}\ \bibnamefont {Oost}}, \bibinfo {author}
  {\bibfnamefont {T.}~\bibnamefont {Klinger}}, , \ and\ \bibinfo {author}
  {\bibfnamefont {R.}~\bibnamefont {Madani}},\ }\bibfield  {title} {\enquote
  {\bibinfo {title} {Probe methods for direct measurements of the plasma
  potential},}\ }\href@noop {} {\bibfield  {journal} {\bibinfo  {journal}
  {Romanian Journal of Physics}\ }\textbf {\bibinfo {volume} {50}},\ \bibinfo
  {pages} {723–739} (\bibinfo {year} {2005})}\BibitemShut {NoStop}%
\bibitem [{\citenamefont {Sheehan}\ \emph {et~al.}(2011)\citenamefont
  {Sheehan}, \citenamefont {Raitses}, \citenamefont {Hershkowitz},
  \citenamefont {Kaganovich},\ and\ \citenamefont
  {Fisch}}]{emissiveprobecomparisionsheehan}%
  \BibitemOpen
  \bibfield  {author} {\bibinfo {author} {\bibfnamefont {J.~P.}\ \bibnamefont
  {Sheehan}}, \bibinfo {author} {\bibfnamefont {Y.}~\bibnamefont {Raitses}},
  \bibinfo {author} {\bibfnamefont {N.}~\bibnamefont {Hershkowitz}}, \bibinfo
  {author} {\bibfnamefont {I.}~\bibnamefont {Kaganovich}}, \ and\ \bibinfo
  {author} {\bibfnamefont {N.~J.}\ \bibnamefont {Fisch}},\ }\bibfield  {title}
  {\enquote {\bibinfo {title} {A comparison of emissive probe techniques for
  electric potential measurements in a complex plasma},}\ }\href@noop {}
  {\bibfield  {journal} {\bibinfo  {journal} {Physics of Plasmas}\ }\textbf
  {\bibinfo {volume} {18}},\ \bibinfo {pages} {073501} (\bibinfo {year}
  {2011})}\BibitemShut {NoStop}%
\bibitem [{\citenamefont {Chen}(2003)}]{chenprobe1}%
  \BibitemOpen
  \bibfield  {author} {\bibinfo {author} {\bibfnamefont {F.~F.}\ \bibnamefont
  {Chen}},\ }\href@noop {} {\enquote {\bibinfo {title} {Lecture notes on
  langmuir probe diagnostics},}\ } (\bibinfo {year} {2003})\BibitemShut
  {NoStop}%
\bibitem [{\citenamefont {Conde}(2011)}]{sphericalprobe2}%
  \BibitemOpen
  \bibfield  {author} {\bibinfo {author} {\bibfnamefont {L.}~\bibnamefont
  {Conde}},\ }\href@noop {} {\enquote {\bibinfo {title} {An introduction of
  langmuir probe diagnostics of plasmas},}\ } (\bibinfo {year}
  {2011})\BibitemShut {NoStop}%
\bibitem [{\citenamefont {Beadles}, \citenamefont {Wang},\ and\ \citenamefont
  {Horányi}(2017)}]{beadlesfloatingpotential1}%
  \BibitemOpen
  \bibfield  {author} {\bibinfo {author} {\bibfnamefont {R.}~\bibnamefont
  {Beadles}}, \bibinfo {author} {\bibfnamefont {X.}~\bibnamefont {Wang}}, \
  and\ \bibinfo {author} {\bibfnamefont {M.}~\bibnamefont {Horányi}},\
  }\bibfield  {title} {\enquote {\bibinfo {title} {Floating potential
  measurements in plasmas: From dust to spacecraft},}\ }\href@noop {}
  {\bibfield  {journal} {\bibinfo  {journal} {Phys. Plasmas}\ }\textbf
  {\bibinfo {volume} {24}},\ \bibinfo {pages} {023701} (\bibinfo {year}
  {2017})}\BibitemShut {NoStop}%
\bibitem [{\citenamefont {Stangeby}(2000)}]{thinsheath}%
  \BibitemOpen
  \bibfield  {author} {\bibinfo {author} {\bibfnamefont {P.~C.}\ \bibnamefont
  {Stangeby}},\ }\href@noop {} {\emph {\bibinfo {title} {The plasma boundary of
  magnetic fusion devices}}}\ (\bibinfo  {publisher} {(Bristol ; Philadelphia,
  PA : Institute of Physics)},\ \bibinfo {year} {2000})\BibitemShut {NoStop}%
\bibitem [{\citenamefont {Delzanno}\ and\ \citenamefont
  {Tang}(2015)}]{chargingsphere}%
  \BibitemOpen
  \bibfield  {author} {\bibinfo {author} {\bibfnamefont {G.~L.}\ \bibnamefont
  {Delzanno}}\ and\ \bibinfo {author} {\bibfnamefont {X.-Z.}\ \bibnamefont
  {Tang}},\ }\bibfield  {title} {\enquote {\bibinfo {title} {Comparison of dust
  charging between orbital-motion-limited theory and particle-in-cell
  simulations},}\ }\href@noop {} {\bibfield  {journal} {\bibinfo  {journal}
  {Physics of Plasmas}\ }\textbf {\bibinfo {volume} {22}},\ \bibinfo {pages}
  {113703} (\bibinfo {year} {2015})}\BibitemShut {NoStop}%
\bibitem [{\citenamefont {Johnson}\ and\ \citenamefont
  {Malter}(1950)}]{doubleprobemalter}%
  \BibitemOpen
  \bibfield  {author} {\bibinfo {author} {\bibfnamefont {E.~O.}\ \bibnamefont
  {Johnson}}\ and\ \bibinfo {author} {\bibfnamefont {L.}~\bibnamefont
  {Malter}},\ }\bibfield  {title} {\enquote {\bibinfo {title} {A floating
  double probe method for measurements in gas discharges},}\ }\href@noop {}
  {\bibfield  {journal} {\bibinfo  {journal} {Phys. Rev.}\ }\textbf {\bibinfo
  {volume} {80}},\ \bibinfo {pages} {58--68} (\bibinfo {year}
  {1950})}\BibitemShut {NoStop}%
\bibitem [{\citenamefont {Nobata}(1963)}]{doubleprobemagnetizedplasma}%
  \BibitemOpen
  \bibfield  {author} {\bibinfo {author} {\bibfnamefont {K.}~\bibnamefont
  {Nobata}},\ }\bibfield  {title} {\enquote {\bibinfo {title} {Characteristics
  of langmuir probe in a strong magnetic field},}\ }\href@noop {} {\bibfield
  {journal} {\bibinfo  {journal} {Japanese Journal of Applied Physics}\
  }\textbf {\bibinfo {volume} {2}},\ \bibinfo {pages} {719} (\bibinfo {year}
  {1963})}\BibitemShut {NoStop}%
\bibitem [{\citenamefont {Tichy}\ \emph {et~al.}()\citenamefont {Tichy},
  \citenamefont {Kudrna}, \citenamefont {Behnke}, \citenamefont {Csambal},\
  and\ \citenamefont {Klagge}}]{trichyprobe2}%
  \BibitemOpen
  \bibfield  {author} {\bibinfo {author} {\bibfnamefont {M.}~\bibnamefont
  {Tichy}}, \bibinfo {author} {\bibfnamefont {P.}~\bibnamefont {Kudrna}},
  \bibinfo {author} {\bibfnamefont {J.}~\bibnamefont {Behnke}}, \bibinfo
  {author} {\bibfnamefont {C.}~\bibnamefont {Csambal}}, \ and\ \bibinfo
  {author} {\bibfnamefont {S.}~\bibnamefont {Klagge}},\ }\bibfield  {title}
  {\enquote {\bibinfo {title} {Langmuir probe diagnostics for medium pressure
  and magnetised low-temperature plasma},}\ }\href@noop {} {\bibfield
  {journal} {\bibinfo  {journal} {Journal de Physique IV Colloque}\ }\textbf
  {\bibinfo {volume} {07 (C4)}},\ \bibinfo {pages}
  {C4--397--C4--411}}\BibitemShut {NoStop}%
\bibitem [{\citenamefont {Kudrna}\ and\ \citenamefont
  {Passoth}(1997)}]{kundraprobe1}%
  \BibitemOpen
  \bibfield  {author} {\bibinfo {author} {\bibfnamefont {P.}~\bibnamefont
  {Kudrna}}\ and\ \bibinfo {author} {\bibfnamefont {E.}~\bibnamefont
  {Passoth}},\ }\bibfield  {title} {\enquote {\bibinfo {title} {Langmuir probe
  diagnostics of a low temperature non-isothermal plasma in a weak magnetic
  field},}\ }\href@noop {} {\bibfield  {journal} {\bibinfo  {journal}
  {Contributions to Plasma Physics}\ }\textbf {\bibinfo {volume} {37}},\
  \bibinfo {pages} {417--429} (\bibinfo {year} {1997})}\BibitemShut {NoStop}%
\bibitem [{\citenamefont {Willis}\ \emph {et~al.}(2012)\citenamefont {Willis},
  \citenamefont {Coppins}, \citenamefont {Bacharis},\ and\ \citenamefont
  {Allen}}]{willisfloatingpotential}%
  \BibitemOpen
  \bibfield  {author} {\bibinfo {author} {\bibfnamefont {C.~T.~N.}\
  \bibnamefont {Willis}}, \bibinfo {author} {\bibfnamefont {M.}~\bibnamefont
  {Coppins}}, \bibinfo {author} {\bibfnamefont {M.}~\bibnamefont {Bacharis}}, \
  and\ \bibinfo {author} {\bibfnamefont {J.~E.}\ \bibnamefont {Allen}},\
  }\bibfield  {title} {\enquote {\bibinfo {title} {Floating potential of large
  dust grains in a collisionless flowing plasma},}\ }\href@noop {} {\bibfield
  {journal} {\bibinfo  {journal} {Phys. Rev. E}\ }\textbf {\bibinfo {volume}
  {85}},\ \bibinfo {pages} {036403} (\bibinfo {year} {2012})}\BibitemShut
  {NoStop}%
\bibitem [{\citenamefont {Benyoucef}\ \emph {et~al.}(2010)\citenamefont
  {Benyoucef}, \citenamefont {Yousfi}, \citenamefont {Belmadani},\ and\
  \citenamefont {Settaouti}}]{collisionacrosssection}%
  \BibitemOpen
  \bibfield  {author} {\bibinfo {author} {\bibfnamefont {D.}~\bibnamefont
  {Benyoucef}}, \bibinfo {author} {\bibfnamefont {M.}~\bibnamefont {Yousfi}},
  \bibinfo {author} {\bibfnamefont {B.}~\bibnamefont {Belmadani}}, \ and\
  \bibinfo {author} {\bibfnamefont {A.}~\bibnamefont {Settaouti}},\ }\bibfield
  {title} {\enquote {\bibinfo {title} {Pic mc using free path for the
  simulation of low-pressure rf discharge in argon},}\ }\href@noop {}
  {\bibfield  {journal} {\bibinfo  {journal} {IEEE Transactions on Plasma
  Science}\ }\textbf {\bibinfo {volume} {38}},\ \bibinfo {pages} {902--908}
  (\bibinfo {year} {2010})}\BibitemShut {NoStop}%
\bibitem [{\citenamefont {Behnke}\ \emph {et~al.}(1999)\citenamefont {Behnke},
  \citenamefont {Passoth}, \citenamefont {Csambal}, \citenamefont {Tich{\'y}},
  \citenamefont {Kudrna}, \citenamefont {Trunec},\ and\ \citenamefont
  {Brablec}}]{Behnkeanisotropyeedf}%
  \BibitemOpen
  \bibfield  {author} {\bibinfo {author} {\bibfnamefont {J.~F.}\ \bibnamefont
  {Behnke}}, \bibinfo {author} {\bibfnamefont {E.}~\bibnamefont {Passoth}},
  \bibinfo {author} {\bibfnamefont {C.}~\bibnamefont {Csambal}}, \bibinfo
  {author} {\bibfnamefont {M.}~\bibnamefont {Tich{\'y}}}, \bibinfo {author}
  {\bibfnamefont {P.}~\bibnamefont {Kudrna}}, \bibinfo {author} {\bibfnamefont
  {D.}~\bibnamefont {Trunec}}, \ and\ \bibinfo {author} {\bibfnamefont
  {A.}~\bibnamefont {Brablec}},\ }\bibfield  {title} {\enquote {\bibinfo
  {title} {A study of the electron energy distribution function in the
  cylindrical magnetron discharge in argon and xenon},}\ }\href@noop {}
  {\bibfield  {journal} {\bibinfo  {journal} {Czechoslovak Journal of Physics}\
  }\textbf {\bibinfo {volume} {49}},\ \bibinfo {pages} {483--498} (\bibinfo
  {year} {1999})}\BibitemShut {NoStop}%
\bibitem [{\citenamefont {Demidov}\ \emph {et~al.}(1999)\citenamefont
  {Demidov}, \citenamefont {Ratynskaia}, \citenamefont {Armstrong},\ and\
  \citenamefont {Rypdal}}]{eedfdiffusion}%
  \BibitemOpen
  \bibfield  {author} {\bibinfo {author} {\bibfnamefont {V.~I.}\ \bibnamefont
  {Demidov}}, \bibinfo {author} {\bibfnamefont {S.~V.}\ \bibnamefont
  {Ratynskaia}}, \bibinfo {author} {\bibfnamefont {R.~J.}\ \bibnamefont
  {Armstrong}}, \ and\ \bibinfo {author} {\bibfnamefont {K.}~\bibnamefont
  {Rypdal}},\ }\bibfield  {title} {\enquote {\bibinfo {title} {Probe
  measurements of electron energy distributions in a strongly magnetized
  low-pressure helium plasma},}\ }\href@noop {} {\bibfield  {journal} {\bibinfo
   {journal} {Physics of Plasmas}\ }\textbf {\bibinfo {volume} {6}},\ \bibinfo
  {pages} {350--358} (\bibinfo {year} {1999})}\BibitemShut {NoStop}%
\bibitem [{\citenamefont {Popov}\ \emph {et~al.}(2012)\citenamefont {Popov},
  \citenamefont {P~Ivanova}, \citenamefont {Kovačič}, \citenamefont
  {Gyergyek},\ and\ \citenamefont {Čerček}}]{eedfsecondderivativeinb}%
  \BibitemOpen
  \bibfield  {author} {\bibinfo {author} {\bibfnamefont {T.~K.}\ \bibnamefont
  {Popov}}, \bibinfo {author} {\bibfnamefont {a.~M.~D.}\ \bibnamefont
  {P~Ivanova}}, \bibinfo {author} {\bibfnamefont {J.}~\bibnamefont
  {Kovačič}}, \bibinfo {author} {\bibfnamefont {T.}~\bibnamefont {Gyergyek}},
  \ and\ \bibinfo {author} {\bibfnamefont {M.}~\bibnamefont {Čerček}},\
  }\bibfield  {title} {\enquote {\bibinfo {title} {Langmuir probe measurements
  of the electron energy distribution function in magnetized gas discharge
  plasmas},}\ }\href@noop {} {\bibfield  {journal} {\bibinfo  {journal} {Plasma
  Sources Science and Technology}\ }\textbf {\bibinfo {volume} {21}},\ \bibinfo
  {pages} {25004--25013} (\bibinfo {year} {2012})}\BibitemShut {NoStop}%
\bibitem [{\citenamefont {Godyak}(1990)}]{godyakeedfbook}%
  \BibitemOpen
  \bibfield  {author} {\bibinfo {author} {\bibfnamefont {V.~A.}\ \bibnamefont
  {Godyak}},\ }\href@noop {} {\emph {\bibinfo {title} {Measuring EEDF in Gas
  Discharge Plasmas. In: Auciello O., Gras-Marti A., Valles-Abarca J.A., Flamm
  D.L. (eds) Plasma-Surface Interactions and Processing of Materials. NATO ASI
  Series (Series E: Applied Sciences), vol 176.}}}\ (\bibinfo  {publisher}
  {Springer, Dordrecht},\ \bibinfo {year} {1990})\BibitemShut {NoStop}%
\bibitem [{\citenamefont {Godyak}, \citenamefont {Piejak},\ and\ \citenamefont
  {Alexandrovich}(1993)}]{godyaknonmaxwellianplasma}%
  \BibitemOpen
  \bibfield  {author} {\bibinfo {author} {\bibfnamefont {V.~A.}\ \bibnamefont
  {Godyak}}, \bibinfo {author} {\bibfnamefont {R.~B.}\ \bibnamefont {Piejak}},
  \ and\ \bibinfo {author} {\bibfnamefont {B.~M.}\ \bibnamefont
  {Alexandrovich}},\ }\bibfield  {title} {\enquote {\bibinfo {title} {Probe
  diagnostics of non‐maxwellian plasmas},}\ }\href@noop {} {\bibfield
  {journal} {\bibinfo  {journal} {Journal of Applied Physics}\ }\textbf
  {\bibinfo {volume} {73}},\ \bibinfo {pages} {3657--3663} (\bibinfo {year}
  {1993})}\BibitemShut {NoStop}%
\bibitem [{\citenamefont {Bohm}, \citenamefont {Burhop},\ and\ \citenamefont
  {Messey}(1994)}]{bohmprobeinmagneticfield}%
  \BibitemOpen
  \bibfield  {author} {\bibinfo {author} {\bibfnamefont {D.}~\bibnamefont
  {Bohm}}, \bibinfo {author} {\bibfnamefont {E.~H.~S.}\ \bibnamefont {Burhop}},
  \ and\ \bibinfo {author} {\bibfnamefont {H.~S.~W.}\ \bibnamefont {Messey}},\
  }\bibfield  {title} {\enquote {\bibinfo {title} {The use of probes for plasma
  exploration in strong magnetic fields},}\ }in\ \href@noop {} {\emph {\bibinfo
  {booktitle} {The Characteristics of Electrical Discharges in Magnetic
  Fields}}},\ \bibinfo {editor} {edited by\ \bibinfo {editor} {\bibfnamefont
  {A.}~\bibnamefont {Guthrie}}\ and\ \bibinfo {editor} {\bibfnamefont {R.~K.}\
  \bibnamefont {Wakerling}}}\ (\bibinfo  {publisher} {McGraw-Hill Book
  Company},\ \bibinfo {address} {New York},\ \bibinfo {year} {1994})\
  Chap.~\bibinfo {chapter} {2}, pp.\ \bibinfo {pages} {13--76}\BibitemShut
  {NoStop}%
\bibitem [{\citenamefont {Sanmartin}(1970)}]{probeinstrongb}%
  \BibitemOpen
  \bibfield  {author} {\bibinfo {author} {\bibfnamefont {J.~R.}\ \bibnamefont
  {Sanmartin}},\ }\bibfield  {title} {\enquote {\bibinfo {title} {Theory of a
  probe in a strong magnetic field},}\ }\href@noop {} {\bibfield  {journal}
  {\bibinfo  {journal} {The Physics of Fluids}\ }\textbf {\bibinfo {volume}
  {13}},\ \bibinfo {pages} {103--116} (\bibinfo {year} {1970})}\BibitemShut
  {NoStop}%
\bibitem [{\citenamefont {Curreli}\ and\ \citenamefont
  {Chen}(2014)}]{chendiffusioncoefficient}%
  \BibitemOpen
  \bibfield  {author} {\bibinfo {author} {\bibfnamefont {D.}~\bibnamefont
  {Curreli}}\ and\ \bibinfo {author} {\bibfnamefont {F.~F.}\ \bibnamefont
  {Chen}},\ }\bibfield  {title} {\enquote {\bibinfo {title} {Cross-field
  diffusion in low-temperature plasma discharges of finite length},}\
  }\href@noop {} {\bibfield  {journal} {\bibinfo  {journal} {Plasma Sources
  Sci. Technol.}\ }\textbf {\bibinfo {volume} {23}},\ \bibinfo {pages} {064001}
  (\bibinfo {year} {2014})}\BibitemShut {NoStop}%
\bibitem [{\citenamefont {Chen}(1984)}]{chenplasmaphysicsbook}%
  \BibitemOpen
  \bibfield  {author} {\bibinfo {author} {\bibfnamefont {F.~F.}\ \bibnamefont
  {Chen}},\ }\href@noop {} {\emph {\bibinfo {title} {Introduction to plasma
  physics and controlled fusion}}}\ (\bibinfo  {publisher} {Springer US},\
  \bibinfo {year} {1984})\BibitemShut {NoStop}%
\bibitem [{\citenamefont {Arnas}, \citenamefont {Mikikian},\ and\ \citenamefont
  {Doveil}(1999)}]{arnasductcharging}%
  \BibitemOpen
  \bibfield  {author} {\bibinfo {author} {\bibfnamefont {C.}~\bibnamefont
  {Arnas}}, \bibinfo {author} {\bibfnamefont {M.}~\bibnamefont {Mikikian}}, \
  and\ \bibinfo {author} {\bibfnamefont {F.}~\bibnamefont {Doveil}},\
  }\bibfield  {title} {\enquote {\bibinfo {title} {High negative charge of a
  dust particle in a hot cathode discharge},}\ }\href@noop {} {\bibfield
  {journal} {\bibinfo  {journal} {Phys. Rev. E}\ }\textbf {\bibinfo {volume}
  {60}},\ \bibinfo {pages} {7420--7425} (\bibinfo {year} {1999})}\BibitemShut
  {NoStop}%
\bibitem [{\citenamefont {Dote}, \citenamefont {Amemiya},\ and\ \citenamefont
  {Ichimiya}(1964)}]{dotecharginginmagneticfield}%
  \BibitemOpen
  \bibfield  {author} {\bibinfo {author} {\bibfnamefont {T.}~\bibnamefont
  {Dote}}, \bibinfo {author} {\bibfnamefont {H.}~\bibnamefont {Amemiya}}, \
  and\ \bibinfo {author} {\bibfnamefont {T.}~\bibnamefont {Ichimiya}},\
  }\bibfield  {title} {\enquote {\bibinfo {title} {Effect of a magnetic field
  upon the saturation electron current of an electrostatic probe},}\
  }\href@noop {} {\bibfield  {journal} {\bibinfo  {journal} {Japanese Journal
  of Applied Physics}\ }\textbf {\bibinfo {volume} {3}},\ \bibinfo {pages}
  {789} (\bibinfo {year} {1964})}\BibitemShut {NoStop}%
\bibitem [{\citenamefont {Khrapak}\ \emph
  {et~al.}(2005{\natexlab{b}})\citenamefont {Khrapak}, \citenamefont
  {Ratynskaia}, \citenamefont {Zobnin}, \citenamefont {Usachev}, \citenamefont
  {Yaroshenko}, \citenamefont {Thoma}, \citenamefont {Kretschmer},
  \citenamefont {H\"ofner}, \citenamefont {Morfill}, \citenamefont {Petrov},\
  and\ \citenamefont {Fortov}}]{kharpakdustneutralcharge}%
  \BibitemOpen
  \bibfield  {author} {\bibinfo {author} {\bibfnamefont {S.~A.}\ \bibnamefont
  {Khrapak}}, \bibinfo {author} {\bibfnamefont {S.~V.}\ \bibnamefont
  {Ratynskaia}}, \bibinfo {author} {\bibfnamefont {A.~V.}\ \bibnamefont
  {Zobnin}}, \bibinfo {author} {\bibfnamefont {A.~D.}\ \bibnamefont {Usachev}},
  \bibinfo {author} {\bibfnamefont {V.~V.}\ \bibnamefont {Yaroshenko}},
  \bibinfo {author} {\bibfnamefont {M.~H.}\ \bibnamefont {Thoma}}, \bibinfo
  {author} {\bibfnamefont {M.}~\bibnamefont {Kretschmer}}, \bibinfo {author}
  {\bibfnamefont {H.}~\bibnamefont {H\"ofner}}, \bibinfo {author}
  {\bibfnamefont {G.~E.}\ \bibnamefont {Morfill}}, \bibinfo {author}
  {\bibfnamefont {O.~F.}\ \bibnamefont {Petrov}}, \ and\ \bibinfo {author}
  {\bibfnamefont {V.~E.}\ \bibnamefont {Fortov}},\ }\bibfield  {title}
  {\enquote {\bibinfo {title} {Particle charge in the bulk of gas
  discharges},}\ }\href@noop {} {\bibfield  {journal} {\bibinfo  {journal}
  {Phys. Rev. E}\ }\textbf {\bibinfo {volume} {72}},\ \bibinfo {pages} {016406}
  (\bibinfo {year} {2005}{\natexlab{b}})}\BibitemShut {NoStop}%
\bibitem [{\citenamefont {Fagan}(2013)}]{magneticfieldlines}%
  \BibitemOpen
  \bibfield  {author} {\bibinfo {author} {\bibfnamefont {M.~A.}\ \bibnamefont
  {Fagan}},\ }\emph {\bibinfo {title} {Fundamental studies of heap leaching
  hydrology using magnetic resonance imaging}},\ \href@noop {} {Ph.D. thesis},\
  \bibinfo  {school} {Department of Chemical Engineering and
  Biotechnology,University of Cambridg} (\bibinfo {year} {2013})\BibitemShut
  {NoStop}%
\end{thebibliography}%
\end{document}